# Germanium Thermophotovoltaic Devices Achieving 7.3% Efficiency Under High-Temperature Emission by Empirical Calorimetry

A. M. Medrano [a*], E. López [a], P. García-Linares [a], J. Villa [a], M. Gamel [b], M. Garín [c], I. Martín [b], C. Cañizo [a], A. Datas [a*]

[a] Instituto de Energía Solar, Universidad Politécnica de Madrid, Av. Complutense s/n, 28040, Madrid, Spain.

[b] Departament d'Enginyeria Electrònica, Universitat Politècnica de Catalunya, C/Jordi Girona 1-3, Mòdul C4, 08034, Barcelona, Spain.

[c] GR-MECAMAT, Department of Engineering, Universitat de Vic – Universitat Central de Catalunya, Vic, Spain

* Corresponding author 1: am.medrano@upm.es

* Corresponding author 2: a.datas@upm.es

**Abstract — We report the first empirical efficiency measurement of germanium-based thermophotovoltaic devices under high-temperature, high-irradiance conditions using a high view-factor calorimetric setup. Two TPV cell architectures were fabricated on p-type, highly doped ($10^{17}$ cm$^{-3}$) Ge substrates, differing only in rear contact configuration. A standard device with a gold rear contact achieves a peak efficiency of 7.3 % and a power density of 1.77 W cm$^{-2}$ at an emitter temperature of 1480 °C, while a PERC-type device reaches 6.3 % efficiency and 1.22 W cm$^{-2}$ at 1426 °C. The superior performance of the standard device is attributed to lower series resistance, whereas the PERC design exhibits slightly higher efficiency at lower emitter temperatures (4.0 % vs. 3.8 % at 1150 °C) due to enhanced rear-surface reflectivity. A detailed TPV model has been developed and validated across both device architectures. The model identifies out-of-band optical losses as the dominant efficiency-limiting mechanism, primarily caused by strong free-carrier absorption in the highly doped Ge substrate. Using this model, we predict device performance under idealized spectral conditions commonly assumed in prior literature. For a simulated AlN/W spectrally selective emitter, efficiencies as high as 22.3 % at 1800 °C are obtained, consistent with previous semi-empirical predictions. In contrast, when previously reported Ge devices are modeled under the realistic graphite emitter spectrum used here, projected efficiencies decrease to as low as 8.1 % at 1480 °C. These results show that earlier projections remain valid but idealized and underscore the importance of emitter spectral engineering and substrate optimization. Finally, we present the first direct comparison of Ge and InGaAs TPV devices under identical conditions, demonstrating the superior performance of InGaAs while confirming the cost-driven competitiveness of Ge.**

## 1 INTRODUCTION

Thermophotovoltaic (TPV) devices convert thermal radiation, primarily in the infrared spectrum, emitted by a high-temperature source into electricity via the photovoltaic effect [1], [2]. This principle makes TPV systems particularly attractive for applications such as industrial waste heat recovery and thermal energy storage [3], [4], especially in the context of renewable energy integration. As renewable energy capacity continues to grow, the need for efficient, dispatchable storage solutions becomes increasingly critical. TPV-based thermal batteries [5], [6] offer a compelling approach for storing excess renewable energy in thermal form and delivering electricity on demand, particularly during periods when generation and consumption are misaligned.

A key distinction between TPV and conventional solar photovoltaic systems lies in the proximity of the device to the heat source [1], [2]. This close coupling benefits from the integration of a reflective mirror placed beneath the cell. This strategy allows unabsorbed radiation to be redirected back to the emitter, enhancing overall system efficiency through radiative recycling. Using this approach, TPV cells based on III-V semiconductor materials have demonstrated efficiencies above 40% [7], [8]. These materials are often employed in advanced structures, such as "air-bridge" cells, which enhance back mirror reflectivity by exploiting the large refractive index contrast between air and metals like gold [7], [9]. However, the use of III-V materials entails high material and processing costs, limiting their industrial scalability [4].

Germanium (Ge) has historically been considered a cost-effective alternative for the fabrication of TPV devices. It has been widely used in the space industry as the bottom cell in multijunction solar cells, benefiting from mature processing techniques and good lattice matching with III-V materials. However, Ge presents specific challenges for TPV applications. For instance, its native oxide does not provide effective surface passivation, and its indirect bandgap inherently limits the open-circuit voltage ($V_{oc}$).

One of the most critical trade-offs in Ge-based TPV design lies in the doping concentration of the substrate. Higher doping levels improve the electrical performance by increasing $V_{oc}$ and lowering the series resistance of the bulk, thus maximizing carrier collection [10], [11]. Nonetheless, elevated doping also increases free-carrier absorption (FCA), particularly for photon energies lower than the bandgap of the absorber, i.e. out-of-band (OOB) photons. These optical losses not only reduce TPV efficiency but also contribute to thermal management issues, as low energy absorbed photons by free carriers are transformed into heat rather than being reflected to the emitter.

Conversely, reducing the doping concentration enhances infrared transparency, allowing a larger fraction of OOB photons to reach the back mirror and be reflected back to the emitter. This boosts optical efficiency but degrades $V_{oc}$ [12]. In addition, the longer diffusion lengths associated with lower doping levels make effective surface passivation increasingly critical. To address this, various strategies have been explored to create passivated contacts at the rear surface of the cells, including the implementation of back-surface field (BSF) structures [12], and the use of passivating dielectric stacks (e.g., $Al_2O_3$/a-Si)[13], [14], [15] for the development of passivated emitter and rear cell (PERC) architectures through the Laser Firing Contact technique (LFC) [13], [16], [17].

These design choices are reflected in the different structures reported in the literature for Ge TPV devices, which span a range of doping concentrations, contact schemes (with their associated active areas and shadowing factors, SF), and rear-side reflectors which deserves a thorough review. A summary of the reported efficiencies of Ge-based TPV devices, along with the main methodological assumptions used for their calculation, is provided in Table 1.

Previous efficiency reports for Ge-based TPV cells are limited and rely exclusively on semi-empirical methods to estimate conversion efficiency under idealized assumptions (see Table 1). In 1963, Wedlock et al. [18] reported a 4.2 % conversion efficiency for a Ge p-i-n device illuminated by a wolfram lamp at 2400 K through a long-pass filter (cutoff at 1 µm), under an incident power density of 282 mW·cm$^{-2}$. This calculation followed a solar-cell-like approach, neglecting sub-bandgap absorption and radiative exchange, unlike modern TPV efficiency definitions. More recently, Fernández et al. [13], [19] reported a 16.5 % efficiency for a LFC-PERC type Ge TPV cell. However, the efficiency calculation assumed a cut-off spectrum using a 1100 ºC microstructured wolfram emitter, effectively neglecting the absorption of OOB photons. Van der Heide et al. [14], [20] do not report TPV efficiencies, but their work provides relevant insights into PERC-type devices with LFC. They investigated the effect of reducing the bulk doping concentration from 1×10$^{17}$ cm$^{-3}$ to 1×10$^{15}$ cm$^{-3}$ and examined the impact of a selective $Er_2O_3$ emitter on short-circuit current and overall device reflectivity under the AM1.5G spectrum for both 1 sun and 20 suns. More recently, Martín et al. [12] reported efficiencies for BSF-type Ge TPV cells fabricated on low (~$10^{15}$ cm$^{-3}$), medium (~$10^{16}$ cm$^{-3}$), and high (~$10^{17}$ cm$^{-3}$) doped substrates, with the highest efficiency (23.2%) obtained for the highest doped case. However, this estimate is also based on a theoretical AlN-coated W spectrally selective emitter that suppressed OOB radiation. In a different approach, Gamel et al. [21] investigated a cost-effective alternative to epitaxial growth for the electron-selective contact, analyzing the use of phosphorus-doped nanocrystalline silicon for the n-type electrode instead of epitaxial layers, combined with an aluminum rear contact. Their model predicts TPV efficiencies of 6.5% and 2.9% for medium (~$10^{16}$ cm$^{-3}$) and low (~$10^{15}$ cm$^{-3}$) doping levels when irradiated with a blackbody emitter at 1328 ºC and 1029 ºC, respectively.

Thus, all prior efficiency values were not obtained through direct calorimetric measurements but were instead derived from models, in most of the cases incorporating idealized assumptions, primarily concerning the spectral distribution of incident radiation on the cell.

*Table 1. Summary of the efficiency reports for Ge-based TPV cells*

| | TPV cell design | | | | | TPV efficiency calculation/measurement conditions | | | | |
|---|---|---|---|---|---|---|---|---|---|---|
| | Substrate doping(s) (cm$^{-3}$) | Front surface | Rear surface | Cell Active Area [cm$^2$] | SF [%] | Emitter | Emitter-cell optical cavity effect | Electric power | Heat dissipated | TPV Efficiency |
| Wedlock et al [18] | p-i-n No dopings reported | p-Ge | n-Ge | 1.29 | - | 1500w Wolfram lamp | Not considered | **Exp. (wolfram lamp)** | Not considered | 4.2 % at 2127 ºC (Simulated) |
| Fernandez et al [13], [19] | $10^{17}$ $10^{16}$ $10^{15}$ | ARC/GaAs/ GaInP | PERC (Si/SiO2 - Al LFC) | 1.55 | - | Sim. (micro-structured wolfram with spectral cut off) | Not considered | **Exp. (under flash lamp)** | Sim. | 16.5 % 1100 ºC (simulated) |
| Martin et al [12] | $10^{17}$ $10^{16}$ $10^{15}$ | GaAs | Al-BSF | 0.09 | 6 | Sim. (AlN/W [22]) | Not considered | **Exp. (under high intensity laser)** | Sim. | 23.2 % ($10^{17}$) 1800 ºC 21.5 % ($10^{16}$) 1800 ºC 20.6 % ($10^{15}$) 1800 ºC (simulated) |
| Gamel et al [21] | $10^{16}$ $10^{15}$ | nc-Si(n) | Al | 0.64 | ~13 | Sim. (Blackbody spectra) | Not considered | **Exp. (under flash lamp)** | Sim | 6.5 % ($10^{16}$) 1328 ºC 2.9 % ($10^{15}$) 1029 ºC (simulated) |
| This work (CONV) | $10^{17}$ | GaAs/ GaInP | Au | 0.85 | 19 | **Exp. (graphite)** | **Exp. (High view factor)** | **Exp. (under high temp emitter)** | **Exp. (calorimetry)** | 7.3 % 1480 ºC (empirical) |
| This work (PERC) | $10^{17}$ | GaAs/ GaInP | PERC (a-SiCx(i)/Al$_2$O$_3$/ a-SiC LFC – Cr/Au) | 0.85 | 19 | **Exp. (graphite)** | **Exp. (High view factor)** | **Exp. (under high temp emitter)** | **Exp. (calorimetry)** | 6.3 % 1425 ºC (empirical) |

This work reports the first fully empirical determination of the efficiency of Ge-based TPV devices. To that end, we developed two types of Ge-based TPV cells: a conventional device (CONV) with a gold rear mirror directly deposited on the Ge substrate, and a PERC-type cell, building upon the concept proposed in [13]. Our design utilizes a highly doped ($10^{17}$ cm$^{-3}$) Ge substrate and incorporates the rear-contact scheme introduced in our earlier work [15], which integrates a multilayer dielectric stack (a-SiCx:H(i)/Al$_2$O$_3$/a-SiC) for effective surface passivation and enhanced rear-side reflectivity, along with localized laser-fired punctual contacts to achieve low contact resistivity and efficient carrier collection.

The devices were characterized using a high-view-factor calorimetric setup equipped with a high-temperature graphite emitter [23], enabling simultaneous, direct measurement of both the electrical power output and the heat dissipated by the cell. This method provides a fully empirical determination of TPV efficiency, in contrast to prior studies that relied on semi-empirical models and idealized assumptions about the spectral distribution of incident radiation.

To get further insight of device performance, we developed a model that accurately reproduces our experimental results. This model is subsequently used to identify and quantify the dominant sources of efficiency loss, most notably, those related to OOB photon absorption in the highly doped Ge substrate. We also applied the model to previously reported device architectures, enabling a direct comparison of device efficiencies and a consistent re-evaluation of their predicted performance under the same assumptions and boundary conditions used in this study.

This article is organized as follows: Section 2 describes the methodology, including TPV cell fabrication, experimental characterization, and numerical simulation. Section 3 presents and discusses the results, while Section 4 compares them with the state of the art for both Ge and InGaAs devices, including the first direct comparison of Ge and InGaAs TPV cells measured under identical experimental conditions.

## 2 METHODOLOGY

### 2.1 MANUFACTURING

In this study, we used 4-inch p-type crystalline Ge wafers, 170 µm thick, oriented along the (100) plane, with a doping concentration of $N_A \approx 2 \times 10^{17} \text{ cm}^{-3}$. Using these substrates, both a baseline Ge TPV cell and a PERC device were fabricated. A front view of the fabricated devices is shown in Figure 1, together with sketches of their respective architectures and a microscope image of the PERC localized rear contacts. A step-by-step visual representation of the manufacturing process can be found in Supplementary information S1.

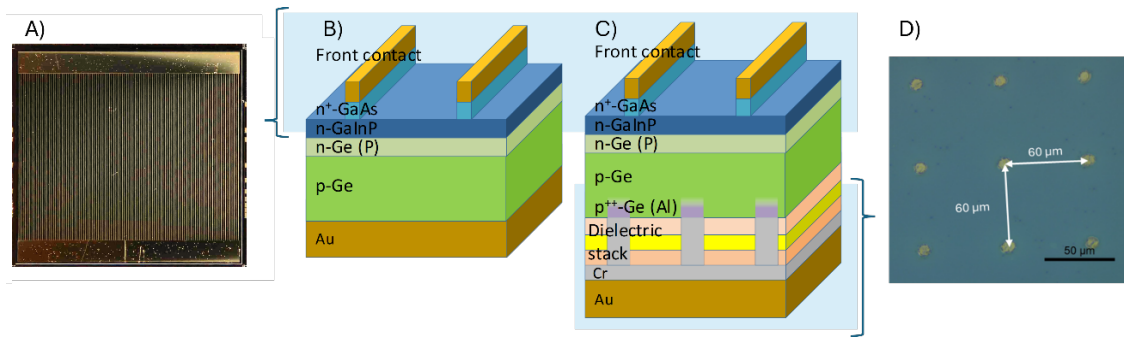

*Figure 1  Structural and visual details of the Ge TPV devices analyzed in this work. A) Front view of the fabricated devices, where the front contact design is identical in both cases. B) and C) Schematic view of the layer configuration for the CONV and PERC respectively (not to scale). D) Microscope image of the localized rear contacts used in the PERC configuration.*

All samples underwent epitaxial growth of n-type GaAs (150 nm)/GaInP (50 nm) layers at the Fraunhofer Institute for Solar Energy Systems (ISE) using metal-organic vapor phase epitaxy (MOVPE). This process simultaneously formed a contact layer for the device emitter, passivated the surface [19], and created the pn junction via phosphorus diffusion in the Ge substrate.

For the front contact, all devices underwent a conventional photolithography followed by thermal evaporation (Joule effect) to deposit an AuGe/Ni/Au metal stack with thicknesses of 85/25/150, nm respectively. This metal scheme is well-established for forming ohmic contacts on n-type GaAs [24], [25] and typically requires a post-deposition annealing step to achieve lower resistivity. However, in our devices, the GaAs contact layer was limited to a thickness of only 150 nm. As a result, annealing the AuGe alloy caused the formation of metal spikes that penetrated through the pn junction, ultimately leading to device short-circuiting. To prevent this, all samples in this work were fabricated without front contact annealing. Consequently, the devices exhibit higher series resistance than expected from fully optimized ohmic contacts. Following the metal evaporation, an additional gold layer was electroplated onto the front contact to increase its height to approximately 5 μm, reducing fingers electrical resistance. The electroplating process, however, proved to be inconsistent, leading to variations in gold resistivity and ultimately contributing to the higher series resistance observed in the PERC device, identified as the main factor behind its lower performance at higher emitter temperature. Before the mesa photolithography step, the GaAs contact layer was selectively removed using an $NH_4OH$ (30 %):$H_2O_2$ (30 %):$H_2O$ solution in a 2:1:10 ratio for 30 s. This step exposed the underlying GaInP while preserving all other layers. Afterward, mesa photolithography and wet etching were performed to electrically isolate the individual devices. The mesa etching consisted of two additional selective steps. First, the GaInP layer was etched using 37 % HCl for 2 s. Second, pn-junction isolation was completed using an $H_3PO_4$ (85 %): $H_2O_2$ (30 %):$H_2O$ mixture in a 1:6:3 ratio for 15 min, reaching depths of up to 5 μm and ensuring electrical isolation between neighbouring devices. All etching steps were conducted at room temperature and did not damage any of the other layers exposed during the process.

For the rear side, all wafers were chemically etched to remove approximately 5 μm of Ge, a critical step to eliminate residual contamination from the previous epitaxial growth which would otherwise result in the formation of an unintended parasitic rear junction, adversely affecting the device behavior. This etch was performed using a solution of $H_3PO_4$ (85 %): $H_2O_2$ (30 %):$H_2O$ (1:6:3) with the front surface protected by photoresist.

For the CONV device, rear metallization was performed by thermally evaporating a 200 nm thick gold layer. In contrast, the PERC devices employed a rear-side dielectric passivation stack designed to enhance both surface passivation and optical reflectivity [15]. This stack consisted of 2 nm of hydrogenated amorphous silicon carbide (a-SiC$_x$:H) deposited by plasma-enhanced chemical vapor deposition (PECVD) at 400 ºC, followed by 50 nm of aluminum oxide ($Al_2O_3$) deposited via atomic layer deposition (ALD), and an additional 45 nm of amorphous silicon carbide (a-SiC) deposited at 370 ºC. The $Al_2O_3$ layer in the stack also serves as a source of aluminum dopants for the subsequent formation of localized LFCs, as described in [15]. Laser firing injects aluminum dopant into the substrate, forming a BSF of p$^+$-Ge that provides low resistivity and highly selective point-like contacts. For this process, a Talon 355-6 Nd:YVO$_4$ laser emitting pulsed ultraviolet light at a wavelength of 355 nm was used. The laser was set to a frequency of 20 kHz, delivering pulses of 9 ns with an emitting power of 11 mW, resulting in a fluence of 0.9 J/cm². The laser spot pattern, determined through previous electrical analysis [15], was configured with a 60 μm pitch and two pulses per spot. Finally, the rear-side metallization was completed by depositing a 5 nm chromium adhesion layer followed by a 200 nm gold layer across the entire rear surface. This step interconnects all rear point-like contacts and simultaneously forms the back surface reflector (BSR).

The final device features an device area of 1.05 cm² (0.85 cm² excluding the busbars), defined by the mesa pattern. A 500 μm separation is left outside the mesa region between adjacent devices to serve as a dicing street and ensure safe wafer cutting.

## 2.2 CHARACTERIZATION

The fabricated cells underwent comprehensive opto-electronic characterization, including optical reflectance measurements over an extended spectral range using Fourier-transform infrared spectroscopy (FTIR) and UV-Vis-NIR spectrophotometry with integrating spheres, external quantum efficiency (EQE) analysis, and both dark and illuminated current–voltage (I–V) characterization. A multi-flash solar simulator I–V measurement system based on a Xe pulsed light, commonly used for concentrated PV (CPV) cell characterization, was employed to evaluate the device performance under high-irradiance conditions [26].

Remarkably, the best-performing cells were characterized using a dedicated high-view-factor calorimetric setup, featuring a high-temperature graphite emitter positioned in the range of 0.7 mm from the front surface of the TPV cell [23]. This extremely close spacing produced an emitter-to-cell view factor close to 0.95, effectively forming an optical cavity that governs the complex radiative exchange between the emitter and the cell [27]. The experimental setup, described in detail in [23], operates in vacuum and includes a tunable high-power continuous-wave diode laser (RFLA500D, 915 nm, up to 500 W) laser for contactless heating of the emitter, as well as a precision calorimeter for measuring the heat dissipated by the cell. To ensure accurate thermal measurements, the TPV cell and the electrical probes (used to measure the output electricity) are maintained at room temperature using two independent cooling systems. This guarantees that all heat generated in the cell is directed through the calorimeter, rather than being dissipated through the electrical contacts. During each measurement, the cell voltage is held at its maximum power point (MPP), allowing the system to reach steady-state conditions. Under these controlled conditions, the TPV conversion efficiency can be directly determined as a function of the measured output power ($P_{el}$) and dissipated heat ($Q_{dis}$) as:

$$\eta_{TPV} = \frac{P_{el}}{P_{abs}} = \frac{P_{el}}{P_{el} + Q_{dis}} \qquad (1)$$

## 2.3 TPV MODEL

A theoretical model was developed to reproduce the performance of the fabricated TPV devices and to gain insight into the underlying loss mechanisms. The goal of this model is to calculate TPV efficiency, as expressed in equation (1), by calculating the following two magnitudes: the net power absorbed by the cell ($P_{abs}$) and the corresponding $P_{el}$.

Calculating $P_{abs}$ involves the computation of the radiative exchange between the emitter and the cell. To that end, we adopt the same approach used in recent literature reporting semi-empirical TPV efficiency values [28] which assumes that both the emitter and the cell are specular surfaces and form an infinitely extended co-planar optical cavity, in which case the absorbed radiative power can be estimated by the following expression [29]:

$$P_{abs} = A_e F_{ec} \int_0^\infty \varepsilon_{eff}(\lambda) \cdot P_{BB}(T_e, \lambda) \cdot d\lambda \qquad (2)$$

where $A_e$ is the emitter area (more precisely, the aperture area that defines the opening through which radiation is transmitted in the experimental setup), $F_{ec}$ is the emitter-to-cell view factor, $P_{BB}$ $(T_e, \lambda)$ is the Plank equation of thermal radiation, which in turn depends on the emitter temperature ($T_e$) and the photon wavelength ($\lambda$), and $\varepsilon_{eff}(\lambda)=\varepsilon_e \cdot \varepsilon_c/(\varepsilon_e+\varepsilon_c-\varepsilon_e \cdot \varepsilon_c)$ is the effective emissivity, which depends on the emitter and cell emissivity, $\varepsilon_e$ and $\varepsilon_c$, respectively. Both $\varepsilon_e$ and $\varepsilon_c$ are experimentally obtained through global reflectivity measurements at room temperature (see Supplementary information S2).

On the other hand, $P_{el}$ is calculated as the product of the current and voltage at the cell MPP. This can be expressed as:

$$P_{el} = I \cdot V|_{\text{MPP}}$$

*(3)*

*I* and *V* are obtained from the well-established single-diode model:

$$I = I_L - I_0 \left( \exp\left[ \frac{q(V + IR_S)}{nk_B T_c} \right] - 1 \right) - \frac{V + IR_S}{R_{sh}}$$

*(4)*

In this equation, $k_b$ is Boltzmann's constant, $T_c$ is the cell temperature, the shunt resistance ($R_{sh}$) is extracted by fitting to dark *I-V* characteristics, i.e., under the condition $I_L = 0$ (See Supplementary information S4). The remaining parameters, ideality factor ($n$), reverse saturation current ($I_0$), and series resistance ($R_s$), are obtained by fitting the model to experimental *I-V* characteristics measured at room temperature under high irradiance conditions, corresponding to different values of photogenerated current ($I_L$) (see Supplementary information S5). The $I_L$, which depends on the $T_e$, is calculated using the following expression [30]:

$$I_L = \frac{qF_{ec}}{hc} * A_e \int_0^\infty \varepsilon_{eff}(\lambda) \cdot P_{BB}(T_e, \lambda) \cdot IQE(\lambda) \cdot d\lambda$$

*(5)*

where *IQE* is the internal quantum efficiency, $q$ is the elementary charge, $h$ is Plank's constant, and $c$ the speed of light. Further information regarding the quantum efficiency measurement is displayed in Supplementary information S2.

The model described by Equations (1) to (5), when provided with the appropriate input device parameters, namely the $T_c$, $F_{ec}$, IQE, $\varepsilon_e$, $\varepsilon_c$, $n$, $I_0$, $R_{sh}$, and $R_s$, can be used to calculate the TPV conversion efficiency as a function of the $T_e$.

The accurate fitting of the model to the high-temperature calorimetry experiments described above require precise knowledge of both the $T_e$ and $T_c$. However, under TPV operating conditions, direct measurement of these temperatures is extremely challenging due to the presence of intense heat fluxes, which lead to significant temperature gradients and thermal non-uniformities across the system. In our setup, temperature sensors were placed as close as possible to the respective surfaces and carefully calibrated. However, despite these precautions, some measurement uncertainties remain. Based on the heat flux direction, the emitter temperature tends to be overestimated, while the cell temperature is typically underestimated.

To address this limitation, a correction procedure is applied to the experimentally measured emitter and cell temperatures. This procedure combines the model described in Equations (1) to (5) with heat conduction equations for both the emitter and the cell, given by the following two expressions:

$$T_e = T_{e,exp} - P_{abs,exp} \cdot r_e \tag{6}$$

$$T_c = T_{c,exp} + Q_{dis,exp} \cdot r_c \tag{7}$$

where $r_e$ and $r_c$ are the effective thermal correction factors related to the thermal resistance of the emitter and the cell, respectively. The procedure involves identifying the values of $r_e$ and $r_c$ that yield corrected emitter and cell temperatures resulting in the best fit between the model and the experimental $I$-$V$ curves. The same values of $r_e$ and $r_c$ are applied across all experiments performed at different emitter temperatures, thereby ensuring that the correction method is physically consistent and independent of specific operating conditions. It is important to note that during the determination of $r_e$ and $r_c$ the saturation current $I_0$ must be also recalculated, through fitting to the $I$-$V$ data, as it depends sensitively on the corrected cell temperature. The results of the model fitting for the best-performing devices presented in this work are provided in Supplementary information S6, including the values obtained for all fitting parameters.

The correction procedure described above, which re-calculates the emitter temperature through model-fitting, is intrinsically sensitive to variations in the model input parameters (IQE, $\varepsilon_e$, $\varepsilon_c$, $n$, $I_0$, $R_{sh}$, $R_s$ and $F_{ec}$). Among them, $F_{ec}$ can be impacted by small variations in the relative positioning of the emitter and the cell. As explained in [23], this uncertainty is confined to approximately 1–2 %, which translates into temperature deviations of only a few degrees in the low-temperature regime and up to about 10 °C at the highest measured temperatures.

## 3 RESULTS AND DISCUSSION

Figure 2 presents the experimentally measured electric power output ($P_{el}$), dissipated heat ($Q_{dis}$), and the resulting TPV efficiency, calculated using Equation (1), as functions of the emitter temperature. These results are shown alongside the corresponding values predicted by the model described above. Data is provided for both CONV and PERC devices, allowing for direct comparison between experimental and modeled results. The detailed numerical values corresponding to these results are summarized in Table S2 and Table S3 of Supplementary information S6.

The results indicate a maximum TPV efficiency of 7.3 % for the CONV device and 6.3 % for the PERC device, with corresponding electrical power outputs of 0.92 W (1.77 W/cm²)[1] and 0.65 W (1.25 W/cm²), respectively. These values were recorded at the maximum emitter temperatures reached during testing: 1480°C for the CONV device and 1426 °C for the PERC device. As the emitter temperature decreases, the efficiency of both devices falls; however, the PERC architecture gains a relative advantage in this regime, for example, achieving 4.0 % at 1150 °C, compared to 3.8 % for the CONV device interpolating the experimental data for the same emitter temperature.

---

[1] In this work, the power density is defined as the $P_{el}$ divided by the emitter area $A_e$, with the emitter taken to be the aperture area of the experimental setup. We use the emitter area rather than the cell area for this calculation because the cell area (1.05 cm²) is considerably larger than the portion that is actually illuminated. Under the high $F_{ec}$ conditions of our setup, the illuminated region of the cell can be well approximated by the aperture area (0.52 cm²). Consequently, reporting power density with respect to the emitter area provides a more accurate measure of the power generated in the actively illuminated region of the cell.

Model predictions are consistent with the experimental observations and indicate that efficiency and power output could be further increased, reaching up to 8.8 % for the CONV device and 6.6 % for the PERC device, with corresponding electrical power outputs of 2.28 W (4.38 W/cm²) and 0.91 W (1.75 W/cm²), respectively, if the emitter temperature were increased to 1816 °C for the CONV device and 1537 °C for the PERC device. These values represent the performance that would be reached assuming the cell temperature remains fixed at the maximum value measured during the experiments (31.0 °C for CONV and 27.7 °C for PERC; See Supplementary information S6).

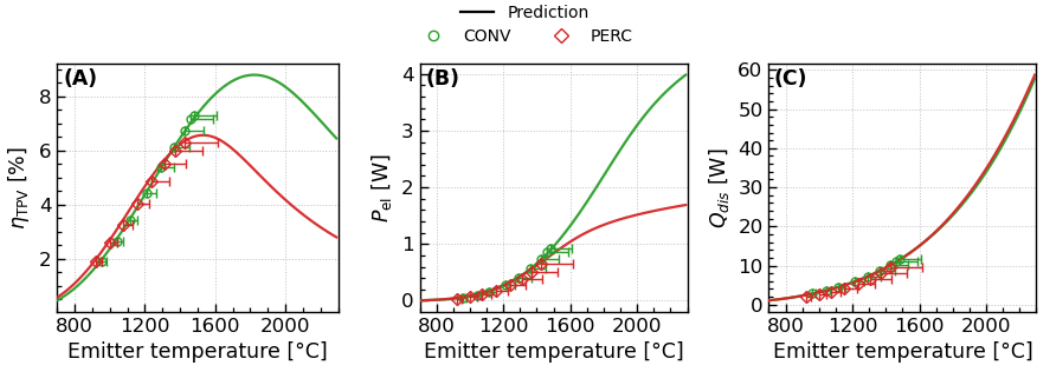

*Figure 2. (A) conversion efficiency $\eta_{TPV}$, (B) electrical power Pel, and (C) heat dissipated Q as functions of the emitter temperature. Error bars in temperature indicate the difference between the corrected temperature (lower bound) and the experimental measurement (higher bound).*

The results reveal significant differences between the CONV and PERC devices. In the high-temperature regime, these differences are mainly driven by variations in series resistance, which becomes increasingly critical at the high current densities generated under elevated irradiance. As a result, the temperature at which maximum efficiency is achieved is reduced for the PERC devices. This interpretation is consistent with the measured resistance values: the CONV device exhibits a series resistance of approximately 7.7 mΩ·cm², while the PERC device reaches around 17.7 mΩ·cm². Two factors account for this disparity. First, methodological: variability in the electroplating process, where increased agitation during film growth improves conductivity in the case of the CONV. Second, structural: point contacts, particularly in highly doped substrates, cannot match the performance of full-area contacts due to the low contact resistance achieved by the later, resulting in a resistance bottleneck on the rear side. A more detailed explanation of these two limiting factors is provided in Supplementary Information S7.

In the low-temperature regime the performance differences between CONV and PERC devices are mainly governed by the slightly higher OOB reflectivity of the PERC architecture, attributable to reduced absorption in the semiconductor/dielectric/metal mirror stack (see Figure S2 in Supplementary information S2). This advantage yields the marginally higher conversion efficiencies observed for PERC devices in this temperature range. Nevertheless, both architectures remain limited by substantial OOB absorption losses, dominated by FCA in the highly doped Ge substrates, which constrains efficiency improvements at lower emitter temperatures.

Figure 3 shows the modelled distribution of absorbed power within the CONV and the PERC TPV cells. These include the $P_{el}$, ohmic losses ($I^2R_s$), and heat absorption losses from both the IB (λ<1.8 μm) and OOB (λ>1.8 μm) spectral ranges. The results of the model confirm that in the low temperature regime, OOB absorption is the dominant loss mechanism, while at higher temperatures IB losses become increasingly significant. Ohmic losses also rise with emitter temperature but remain relatively minor, between 1

% and 2 % (absolute) of the total losses at the maximum measured efficiency point (~1470 ºC) for the CONV and PERC cells, respectively.

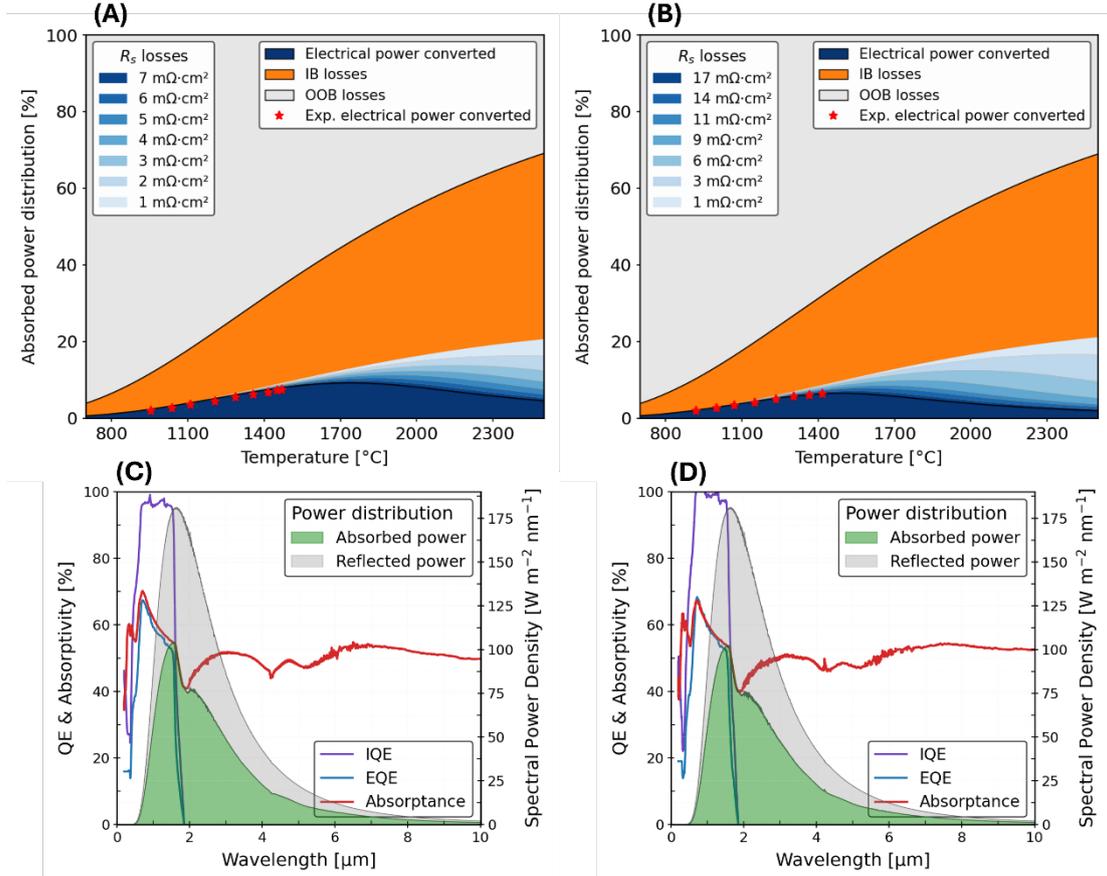

*Figure 3. Temperature-dependent absorbed power distributions for the CONV (A) and PERC (B) devices. The incident power is partitioned into absorbed power (IB and OOB), electrical power converted, and then the IB partition related to ohmic losses is represented for different series resistance values. Red stars indicate the experimentally measured electrical power conversion of the total power absorbed (TPV efficiency). Panels (C) and (D) show the spectral power emitted by the graphite emitter at 1480 °C, split into absorbed and reflected contributions for the CONV and PERC devices, respectively. The wavelength-dependent EQE, IQE, and absorptance are overlaid to relate the spectral absorption with the carrier collection efficiency.*

# 4    COMPARISON WITH STATE-OF-THE-ART RESULTS

## 4.1    GE TPV CELLS: SEMI-EMPIRICAL PREDICTIONS

The state of the art, as previously discussed, explores several strategies to enhance the performance of Ge-based TPV devices, many of which rely on simulated spectrally selective emitters. Among these works, the only publication that provides a complete experimental dataset enabling a rigorous comparison with our model is that of Martin et al. [12], which also reports the highest efficiency obtained using a semi-empirical approach. Other studies either employ low-doped Ge substrates (e.g., Mansur et al.[21]) or lack key information necessary for accurate simulation, as in the case of Fernández et al [13]. For context, Supplementary Information S8 analyses in detail the efficiency-calculation methodologies used in the only two reported TPV devices based on highly doped Ge substrates, those of Martin et al. [12] and Fernández et al. [13].

Using a semi-empirical approach, Martín et al. [12] reported TPV efficiencies of 23.2 % at an emitter temperature of 1800 °C when pairing a Ge TPV cell (also fabricated on a highly doped substrate with a doping concentration of $10^{17}$ cm$^{-3}$) with a theoretical spectrally selective Al/W emitter [22]. Assuming a spectrally selective emitter is the most straightforward strategy to improve the efficiency of Ge TPV cells fabricated on thick, highly doped substrates, as this approach provides significantly larger gains than the use of BSR, whose impact is inherently limited by the strong OOB FCA of the Ge substrate.

The cells reported in [12] are theoretically capable of operating at very high emitter temperatures due to an exceptionally low series resistance of 1 mΩ cm² strongly favored by a small cell area (0.36x0.36 cm²) and the use of a 4-busbar front-electrode configuration. In addition, the devices exhibited higher EQE (see Figure S5 in Supplementary information S3), primarily due to a lower shadowing factor (6 %). This advantage likewise stems from the smaller cell size, which eliminates the need for a dense front metal grid to efficiently collect current.

By contrast, the devices developed in this work utilize a larger cell area (1.05 cm²), a simpler 2-busbar front-contact layout, and a non-annealed front contact (see Supplementary information S7). These design choices result in higher series resistance (7.7 to 17.7 mΩ·cm²) and reduced performance at elevated temperatures. The larger area also necessitates a denser front grid, leading to increased shadowing factor (19 %) and consequently lower EQE.

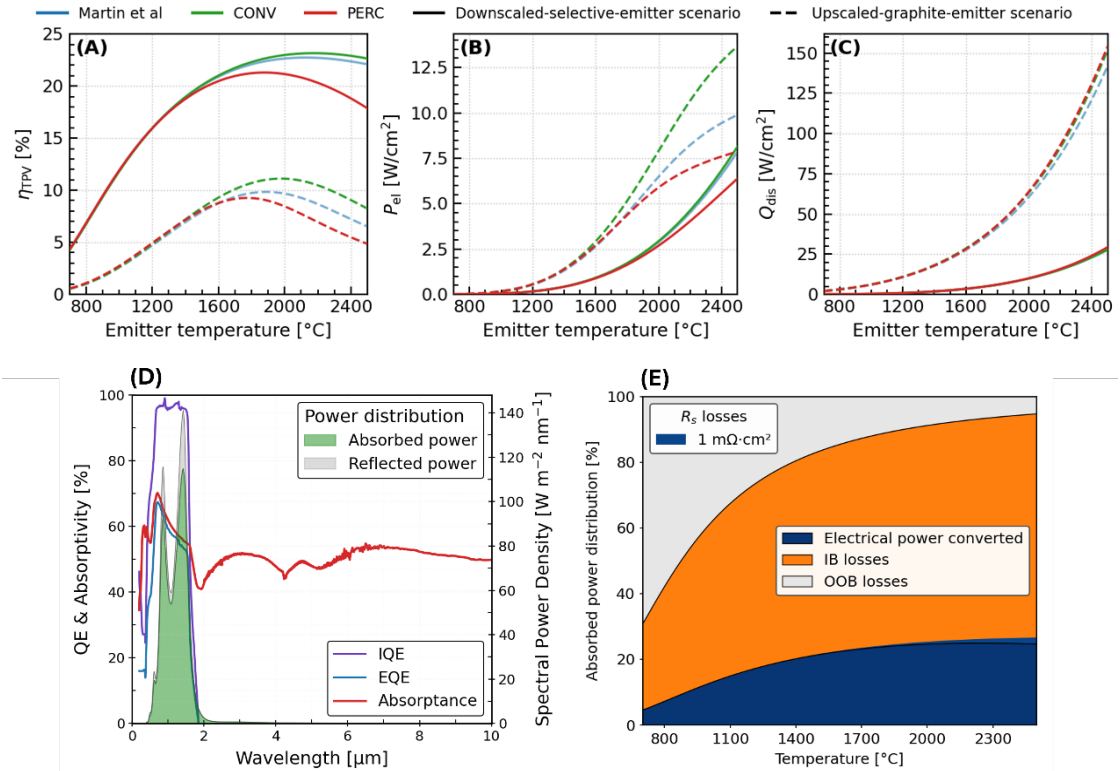

*Figure 4. Simulations of the TPV converters presented in this work and by Martin et al. The data for efficiency (A), power density (B), and dissipated heat density (C) are shown for two scenarios: (1) devices with an active area of 0.52 cm² illuminated by a graphite emitter (dashed lines), and (2) devices with an active area of 0.09 cm², as reported by Martin et al.[12], illuminated by a spectrally selective AlN/W emitter (solid lines). Panel (D) shows the spectral distribution of absorbed and reflected power for the CONV device under illumination from an AlN/W emitter at 1800 °C, together with the corresponding wavelength-dependent IQE, EQE, and absorptance. Panel (E) presents the temperature-dependent absorbed power distribution for the CONV device, separating the contributions from absorbed power (IB and OOB), electrical power converted, and ohmic losses related to the IB loses.*

Figure 4 shows the predicted performance of the CONV and PERC cells developed in this work in the *Downscaled–selective-emitter* scenario where our devices are adjusted to the dimensions of the devices reported in [12] (0.09 cm²) and illuminated by the same theoretical spectrally selective emitter described in [22] (solid lines). Conversely, it also presents the *Upscaled–graphite-emitter* scenario, where the devices from [12] are adjusted to the dimensions of our cells and irradiated by the experimental graphite emitter used in this study (dashed lines). This rescaling includes corrections for geometrical shadowing, with the corresponding optical adjustments, as well as a correction of the series resistance to match the front-grid design of our devices. The procedure for $R_s$ rescaling is detailed in Supplementary information S7, while the optical and shadowing adjustments follow equations provided in S2 and S3. Specifically, the values used are consistent across both studies: for the large-cell configuration (1.05 cm², 15 µm finger width, 100 µm spacing, SF = 19 %), the series resistances are 3.3 mΩ cm² (CONV), 6.2 mΩ cm² (PERC), and 4.7 mΩ cm² (Martín et al [12]). For the small-cell configuration (0.09 cm², 6 µm finger width, 94 µm spacing, SF ≈6 %), the corresponding $R_s$ values are 1.4 mΩ cm² (CONV), 4.3 mΩ cm² (PERC), and 1.1 mΩ cm² (Martín et al [12]).

Remarkably, the results calculated in [12] do not consider the emitter-cell optical cavity effects, accounted for by the effective emissivity in equation (2). To enable a fair comparison, we first reproduced the results in [12] without considering emitter-cell cavity effects. We then re-evaluated the same cases with our simulation framework, which explicitly includes emitter–cell interactions, with cell temperature fixed at 25 °C, and computed the efficiencies shown in Figure 4.

Noteworthy, the simulated results in Figure 4 for both CONV and PERC devices in the *Upscaled–graphite-emitter* scenario are greater than the experimental results shown in Figure 2. This is because the simulations assume improved series resistance (annealed contacts), as discussed in Supplementary information 7.

In the *Downscaled–selective-emitter* scenario shown in Figure 4A our model predicts TPV efficiencies of 22.3 % (CONV), 21.2% (PERC), and 22.1% (Martin et al) at 1800 °C (the temperature reported in [12]). These are not the operating maxima: extending the temperature range yields peaks of 23.2 % at 2151 °C for CONV, 21.8 % at 1858 °C for PERC, and 22.7 % at 2123 °C for the device of Martin et al. Notably, the efficiency of 22.1 % predicted at 1800 °C for the Martin et al. device is lower than the value reported in their study. This discrepancy arises because our calculations account for the emitter–cell optical cavity, an effect that was neglected in their analysis.

Conversely, in the *Upscaled–graphite-emitter* scenario, the cells reported in [12] would achieve an efficiency of 8.1 % at an emitter temperature of 1480 °C the maximum temperature register in our experiments, closely matching the experimental observations obtained with our cells. Under the same graphite-emitter conditions and accounting for the potential series-resistance improvements discussed in Supplementary information 7 (i.e. annealed contacts), our devices would reach 8.1 % (CONV) and 7.9 % (PERC) at 1480 °C. As in the previous case, the temperature of maximum efficiency is higher: the achievable peaks are 11.1 % at 1983 °C for CONV, 9.3 % at 1774 °C for PERC, and 9.8 % at 1858 °C for the device of Martin et al. [12].

Two conclusions can be extracted from the comparison presented above. On the one hand, the use of spectrally selective emitters plays a crucial role in TPV device performance, especially when parasitic absorption of OOB is the dominant loss mechanism. On the other hand, the dramatical difference between experimental efficiency values and idealized semi-empirical projections based on tailored emitter spectra highlights the difficulty of standardizing efficiency calculations in TPV systems. Semi-empirical models, such as the one presented in [12], often rely on idealizations that fail to reflect the performance of real devices. Furthermore, their accuracy depends on precise knowledge of all relevant optical, thermal and electrical properties of constituent materials, as well as their mutual interaction. This requirement is at odds with the intrinsic complexity and variety of TPV systems, including factors such as emitter spectral characteristics, view factor, optical cavity design, and device geometry. The results reported in this work therefore emphasize the importance of direct experimental calorimetry measurements which, when rigorously implemented, provide a reliable basis for assessing TPV performance under realistic operating conditions.

## 4.2 INGAAS TPV CELLS: DIRECT CALORIMETRY MEASUREMENTS

Experimental efficiency measurements of TPV devices based on InGaAs PV absorbers have been widely reported in the recent literature, providing a useful reference point for placing the performance of Ge-based converters into context. Among these studies, only a subset employs direct calorimetry, i.e., measurements in which both $Q_{dis}$ and $P_{el}$ are experimentally determined, allowing a true in-situ determination of TPV efficiency. To establish a consistent basis for discussion, Table 2 compiles the key experimental parameters reported in studies that quantify TPV efficiency using calorimetric techniques. To better visualize the performance trends across these studies, the compiled data are also represented in Figure 5.

Among these studies, our previous work (López et al. [23]) was carried out using the same calorimetric setup as in the present study. This unique overlap enables the first fully consistent comparison between InGaAs and Ge TPV devices under identical experimental conditions, as discussed later in this section.

*Table 2 Summary of experimental reports on TPV efficiencies obtained through calorimetry methods.*

| Work | Material | Cell Area/ Aperture Area[i] [cm²] | Emitter material | Emitter-to-Cell view factor $F_{ec}$ | $P_{el}$ [W] @$T_{emitter, max}$ (Power density [W/cm²]) | Max TPV efficiency reported |
|---|---|---|---|---|---|---|
| Wernsman et al [31] | InGaAs | 4.0/4.0 | SiC ion-etched | - | 3.16 (0.79)[iv] | 23.6 @ 1039 °C |
| Siergiej et al [32] | InGaAs/InPAs | 4.0/4.0 | SiC ion-etched | - | 3.60 (0.9)[ii] | 16.5 @ 1058 °C |
| Lopez et al [23] | InGaAs | 1.05/0.52 | Graphite | 0.95 | 2.51 (4.85)[v] | 26.4 % @ 1592 °C |
| Mahorter et al [33] | InGaAs/InPAs | 4.0/4.0 | SiC ion-etched | - | - | 20.6 % @ 1060 °C |
| Narayan et al [34] | GaAs InGaAs | 0.91/0.91 0.62/0.62 | Graphite | - | 2.23 (2.45)[ii] 0.41 (0.66)[ii] | 31 % @ 2330 °C 30 % @ 1300 °C |
| Tervo et al [35] | GaInAs GaInPAs/ GaInAs | 0.8/0.64 | Graphite | 0.31 | 2.42-3.02 (3.78)[iii] 3.70-4.62 (5.78)[iii] | 38.8% @ 1850 °C 36.8 % @ 1900 °C |
| LaPotin et al [8] | InGaAs multijunction 1.4/1.2 1.2/1.0 | 0.81/ 0.72 | Wolfram halogen bulb | 0.107 | 1.72-1.94 (2.39)[iii] 1.30-1.46 (1.80)[iii] | 41.1% @ 2400 °C 38.9 %@ 2178 °C |
| CONV | Ge | 1.05/0.52 | Graphite | 0.95 | 0.92 (1.77)[iv] | 7.3 % @ 1485 °C |
| PERC | Ge | 1.05/0.52 | Graphite | 0.95 | 0.65 (1.22)[iv] | 6.3 % @ 1432 °C |

[i] *The cell area corresponds to the total mesa-defined cell area (i.e., the area enclosed by the mesa etch), while the aperture area defines the opening through which radiation is transmitted. Owing to beam*

*divergence downstream of the aperture, the illuminated region on the cell may extend beyond the aperture footprint.*

ii *The study reports power density only. Total power was calculated by multiplying the reported power density by the aperture (cell) area, which are identical in this work.*

iii *The study reports power density only and does not specify the reference area. The power range shown was obtained by multiplying the reported power density by the aperture area (lower bound) and the cell area (upper bound).*

iv *The study reports both power and power density, with power density defined as total power divided by the aperture area.*

v *The study reports both absolute power and power density. Power density is defined as the total electrical power normalized between two possible areas: the graphite emitter area ($0.5 \ cm^2$) or to the TPV cell active area ($0.7 \ cm^2$). For consistency throughout this work, all power-density calculations are normalized to the aperture area of $0.52 \ cm^2$, as detailed in the footnote on page 8 [23].*

Figure 5A shows electrical power as a function of dissipated heat, while Figure 5B shows the electrical power density as a function of total dissipated heat density, including results from [8], [23], [31], [32], [33], [34], [35]. Each point in these graphs corresponds to a unique heat-power combination, with the corresponding TPV efficiency indicated by the green color scale in the background. The marker color, following a red-scale colormap, corresponds to the emitter temperature at which the device was tested. Figure 5C and D show the electrical power density and conversion efficiency as a function of the emitter temperature for the same experimental results shown in Figure 5A and B.

The figure highlights the markedly higher efficiency of InGaAs-based devices, founded on their thin-film architecture due to the very high absorption coefficients inherent in direct-bandgap semiconductors. This design supports higher $V_{oc}$, lower $R_s$, and superior OOB photon reflection. By contrast, Ge-based TPV devices suffer from substantial OOB absorption due to their thick substrates, which also limit the achievable $V_{oc}$. As a result, Ge-based devices exhibit not only lower efficiency but also reduced power density, driven mainly by the lower $V_{oc}$ and short-circuit current.

It is important to note that the reduced power density of Ge-based devices is not directly evident in Figure 5C, since the experimental studies compared there employed different view factors and thermal emitters. These variations explain the discrepancies in absolute power and power density across the datasets.

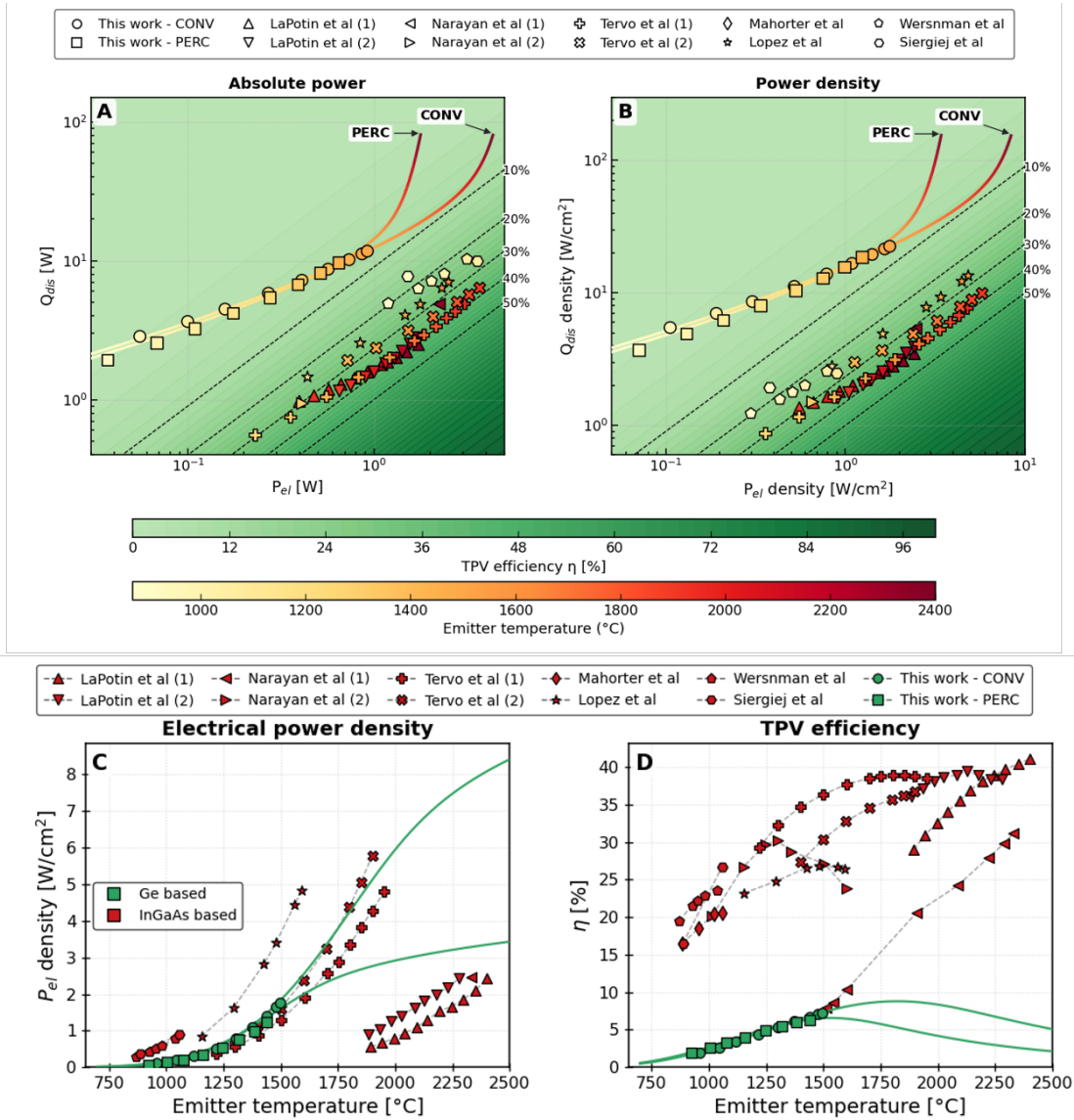

Figure 5. Measured (A) heat and power and (B) heat and power densities, for state of the art InGaAs TPV devices, measured using calorimetric techniques, as reported in Table 2. Each data point in graphs (A) and (B) corresponds to a unique heat-power combination, with the corresponding TPV efficiency indicated by the green color scale. The marker color, following a red-scale colormap, corresponds to the emitter temperature at which the device was tested. (C) shows the power density as a function of emitter temperature, based on calorimetric measurements normalized by the illuminated area. (D) displays TPV efficiencies obtained through calorimetric methods as a function of emitter temperature. Some data points are absent from specific panels due to missing information required for plotting, such as incomplete calorimetric power reporting (e.g., Mahorter et al. [33]). For Tervo et al. [35] and Lapotin et al. [8], the absolute electrical power in panel (A) was calculated assuming an average between the aperture area and the total cell area, because the original works do not clearly state which area was used to compute the reported power.

Given the methodological equivalence highlighted above, Figure 6 presents experimental results for InGaAs devices reported in our earlier work [23] alongside those of the CONV cell examined here. All results are referenced to the emitter temperature directly measured in our setup, without applying the thermal emitter temperature correction described above. Such a correction cannot be applied to the results in [23], as the necessary optical characterization data are not available.

As shown in Figure 6, under identical experimental conditions (i.e. an uncorrected emitter temperature of approximately 1590ºC), InGaAs devices achieved an efficiency of 26.4 % and an output power of 2.51 W ($V_{mpp}$= 416 mV, $I_{mpp}$ = 6.04 A), whereas Ge TPV cells reached only 7.1 % efficiency with an output power of 0.86 W ($V_{mpp}$ = 285 mV, $I_{mpp}$ =

3.0 A). These results demonstrate that InGaAs TPV cells deliver nearly three times the output power and approximately 3.5 times the conversion efficiency of Ge cells under the same conditions. This direct comparison highlights the clear performance advantage of InGaAs. Noteworthy, this advantage would likely be even greater if record-performing InGaAs devices with enhanced OOB photon recycling and improved carrier collection were considered [8], [30], [35].

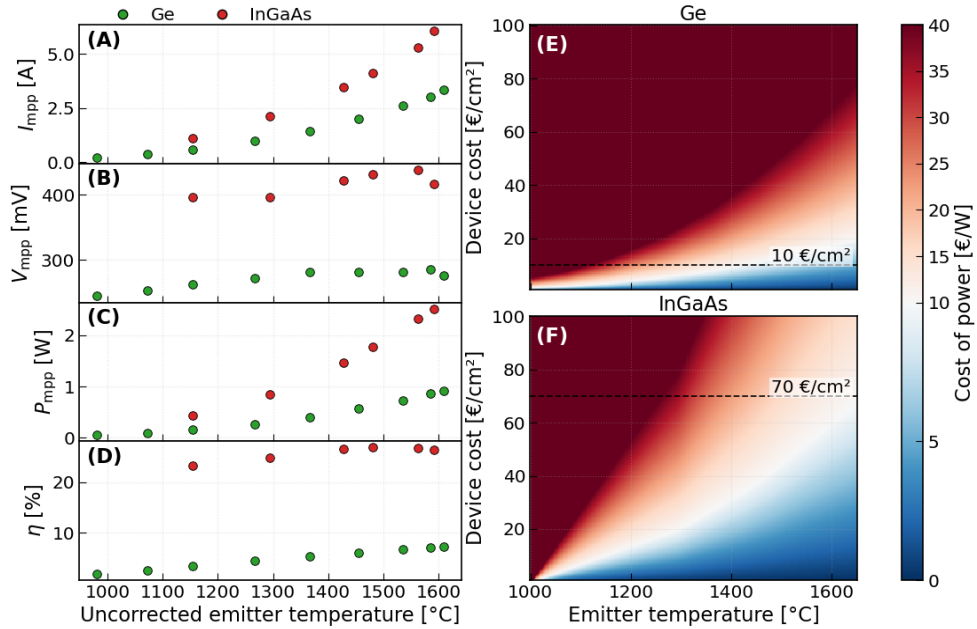

*Figure 6. Cost and performance comparison of Ge and InGaAs devices as a function of the uncorrected emitter temperature (°C). Panels show (A) $I_{mpp}$ [A], (B) $V_{mpp}$ [mV], (C) $P_{mpp}$ [W], and (D) η[%]. Panels (E) and (F) present the cost of power (in €/W) as a function of emitter temperature (in °C) and the cost per unit of cell area (in €/cm²) for both Ge and InGaAs devices, respectively. The cost of power (€/W) is computed as the ratio of the cost per area (in €/cm2) and the measured electrical power density (in W/cm²). The color scale highlights regions with low cost of power (blue, < 10 €/W) and high cost of power (red, > 10 €/W). The dashed horizontal lines indicate representative device costs used for comparison (10 €/cm² for Ge and 70 €/cm² for InGaAs [4]), allowing direct reading of the resulting €/W versus emitter temperature at those assumed cost levels.*

Nonetheless, while InGaAs-based devices clearly outperform Ge in both efficiency and power density, Ge retains a competitive niche in applications where the cost of power (€/W) is the dominant techno-economic metric, such as industrial waste-heat recovery. In these scenarios, the low operating cost of TPV cells—driven by the low cost of the incident photons [4] – means that the levelized cost of electricity is largely determined by the capital expenditure of the TPV converter itself (in €/W), rather than by its efficiency [3], [4].

The lower material and fabrication cost of Ge can therefore compensate for its reduced electrical performance. For example, at an emitter temperature of 1455 °C, a Ge TPV cell with a device cost of 10 €/cm² [4] and a power density of 1.09 W/cm² results in a cost of power of 9.2 €/W. Under the same conditions, an InGaAs device costing 70 €/cm² [4] and delivering 3.41 W/cm² yields 20.5 €/W, corresponding to a cost ratio of approximately 2.2 in favor of Ge. Similar trends are obtained when the analysis is extended across the investigated emitter temperature range, with the Ge-to-InGaAs cost-of-power ratio varying between approximately 2.2 and 2.5 for the cost assumptions considered. Further insights are provided in Figure 6, panels E and F, which illustrates the cost of power as a function of both emitter temperature and device cost for both Ge (panel E) and InGaAs (panel F) devices.

By contrast, the use of higher efficiency InGaAs devices becomes critical in applications with substantial operational expenditures, that is, where the incident photons carry a high associated cost, such as thermal batteries or fuel-driven TPV systems. In these cases, increased TPV efficiency directly reduces the thermal input required per unit of electrical output, thereby lowering operating costs. Under such conditions, the superior efficiency of InGaAs-based TPV cells can justify their higher capital cost, particularly at sufficiently high emitter temperatures where high power densities can be sustained [4].

The use of lower-efficiency but low-cost Ge devices can also be justified in applications with significant operational costs (e.g., thermal batteries or fuel-driven TPV systems) when the heat dissipated in the cell is recovered for heating purposes. In such cases, leveraging this waste heat can increase the overall system efficiency, enabling economically attractive combined heat and power solutions for buildings and low-temperature industrial applications [36].

## 5 CONCLUSIONS

In this work we have provided, to our knowledge, the first fully empirical determination of Ge-based TPV efficiencies under high-temperature, high-irradiance operation. Two device architectures on highly doped p-type Ge ($\approx 10^{17}$ cm$^{-3}$) were fabricated and tested under these conditions, using a graphite emitter and a high view-factor ($\approx 0.95$) calorimetric setup. The conventional design with a full-area Au rear contact achieved 7.3 % efficiency and 1.77 W cm$^{-2}$ at 1480 °C, while the PERC-type device reached 6.3 % and 1.22 W cm$^{-2}$ at 1426 °C. The higher peak efficiency of the conventional cell is attributed to its lower series resistance under the large current densities induced at elevated temperatures. Conversely, at lower emitter temperatures the PERC architecture exhibits a modest advantage (e.g., 4.0 % vs. 3.8 % at 1150 °C) due to improved rear-side reflectivity that curtails sub-bandgap absorption.

A theoretical model, validated against both architectures across temperature, reproduces the measurements and clarifies the internal loss budget. The analysis identifies OOB photon losses, dominated by FCA in the highly doped Ge substrate, as the primary efficiency limiter under the experimental (graphite emitter) conditions. Using the validated model, we reconciled prior literature by re-evaluating state-of-the-art devices under two scenarios. On the one hand, devices in the literature [12] modeled under our graphite spectrum, results in efficiencies ~8.1 % near 1450–1500 °C, consistent with the efficiencies experimentally measured for our devices. On the other hand, under idealized, spectrally selective AlN/W emission, our own conventional device could approach efficiencies of approximately 22 % at temperatures around 2000 ºC, in line with earlier semi-empirical projections.

Finally, this work also provided the first direct comparison between Ge- and InGaAs-based TPV devices measured under an identical experimental conditions, including the same emitter, view factor, and calorimetric protocol, thereby avoiding ambiguities associated with purely electrical or optically inferred metrics. Under these conditions, InGaAs devices achieve approximately three times higher output power and conversion efficiency than Ge cells. This difference is associated with the thin-film architecture of InGaAs, which enables higher operating voltages, reduced series resistance, and more effective OOB photon recycling. In contrast, thick, highly doped Ge substrates exhibit stronger free-carrier absorption and enhanced non-radiative recombination, resulting in lower operating voltages as well as reduced efficiency and power density.

Despite its lower performance, the lower cost of Ge TPV cells represents a clear advantage in applications with low operational costs, such as industrial waste-heat recovery, where the capital expenditure becomes the dominant cost component. Under these conditions, Ge TPV cells must be approximately two to three times less expensive than InGaAs cells to achieve a lower cost per unit power capacity —a condition that is currently met in practice. Conversely, in applications where the thermal input carries a significant cost, such as thermal batteries or fuel-driven TPV systems, operational expenditures become increasingly important, and higher TPV efficiency plays a more significant role in reducing overall levelized cost of generated electricity.

Future work should focus on improving the performance of germanium-based TPV devices through effective spectral control strategies, such as spectrally selective emitters or optical filters, to better match the incident spectrum to the Ge bandgap. In addition, substrate optimization—particularly through the use of lower-doped and/or thinner substrates—can reduce free-carrier absorption and non-radiative recombination losses. Together, these approaches offer a clear pathway to enhancing efficiency and power density, further improving the viability of Ge TPV cells in a broader range of applications.

## ACKNOWLEDGEMENTS

The *THERMOBAT* project received funding from the European Commission under grant agreement No. 10105754. The sole responsibility for the content of this publication lies with the authors. It does not necessarily reflect the views of the European Union. Neither the REA nor the European Commission is responsible for any use that may be made of the information contained herein.

The authors also acknowledge financial support from the Spanish Research State Agency (MCIN/AEI/10.13039/501100011033) through the projects PID2020-113533RB-C31 (GREASE), PID2020-115719RB-C21 and PID2020-115719RB-C22 (GETPV), TED2021-131778B (TROPIC), and PID2023-150209OB-C21 and PID2023-150209OB-C22 (TOMATE). Project TED2021-131778B (TROPIC) is additionally co-funded by the European Union through the "NextGenerationEU" recovery plan and PRTR.

The authors wish to thank Dr. David Lackner and Dr. Frank Dimroth (Fraunhofer ISE) for the development of the epitaxial structures used in the fabrication of the TPV cells. They also gratefully acknowledge Pablo Martín for generously sharing experimental data from his devices, which enabled a thorough and accurate comparative analysis. The authors further thank Dr. Iván García and Rubén Fortín for performing the EQE measurements on multiple devices. Special thanks are extended to Dr. Rebeca Herrero and the ISI group at IES-UPM for providing access to the HELIOS 3030 flash tester and for their support during the measurements.

**Supplementary information**

### S1. Fabrication procedure

Figure S1 provides a comprehensive illustration of the methodology employed in the fabrication of the devices analyzed in this study, highlighting the key steps and materials involved throughout the process.

Conventional (CONV) devices were manufactured following the next sequence of steps: 1) front electrode metallization by thermal evaporation of an AuGe/Ni/Au stack with respective thicknesses of 85 nm / 25 nm / 100 nm, followed by gold electroplating up to approximately 5 μm; 2) mesa etching of the device to remove the contact layer and define the active area; and 3) rear metallization by deposition of a 200 nm-thick gold layer forming the rear electrode and back mirror.

Passivated Emitter Rear Contact (PERC) devices were manufactured following the next sequence of steps: 1) rear passivation was performed as described in the main text, consisting of the deposition of an a-SiCx:H (2 nm) / Al$_2$O$_3$ (50 nm) / a-SiC (45 nm) stack; 2) front electrode metallization identical to that of the CONV device; 3) localized rear contacts were created through the dielectric stack using a Laser-Fired Contacts (LFC) process; 4) a Cr/Au (5 nm / 200 nm) bilayer was deposited to complete the rear electrode; and 5) a final mesa etch defined the device area and removed the contact layer outside the active region. Further details can be found in the Methodology section in the main body of this work.

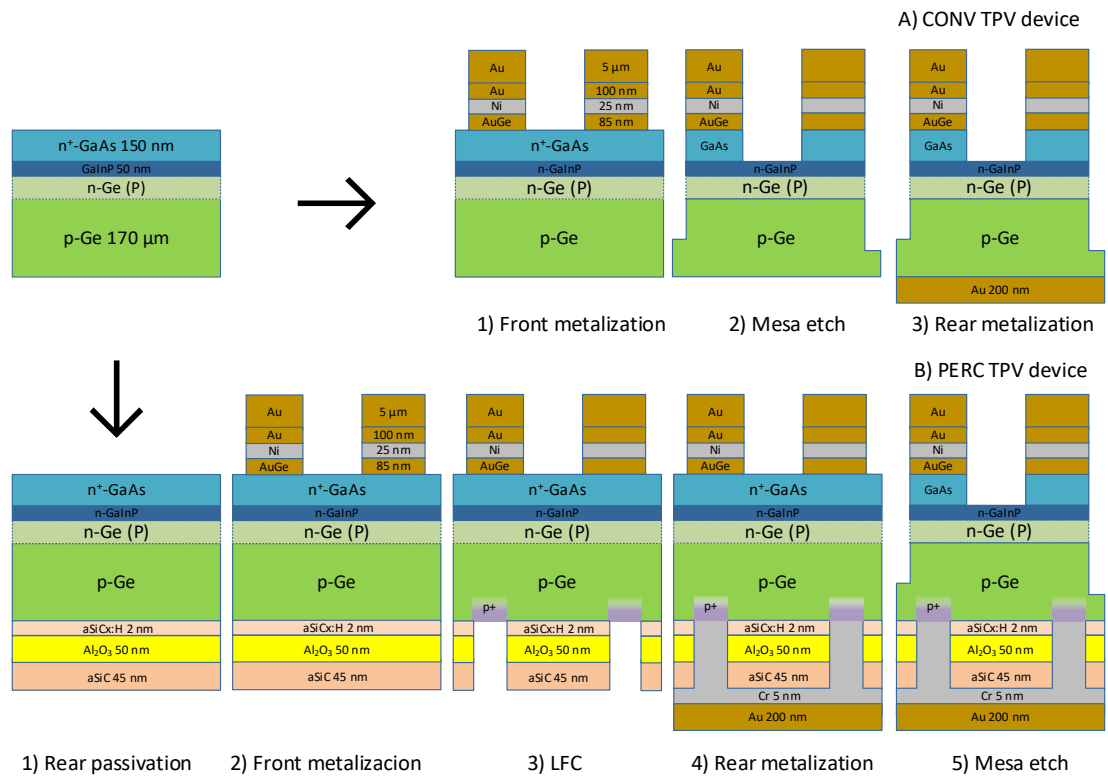

*Figure S1 Manufacturing process step by step for both CONV (Panel A) and PERC (panel B) Ge TPV devices.*

### S2. Spectral properties of TPV cells and emitters

The optical reflectivity ($R_\lambda$) of the emitter and the TPV cells has been measured at room temperature across the relevant wavelength range using two complementary systems: 1) a Nicolet 6700 FTIR equipped with a PIKE integrating sphere for the infrared range (2–15 µm), and 2) a Cary 7000 spectrophotometer with a built-in integrating sphere and a multiple-detector setup for the UV–Vis–NIR range (0.2–2.5 µm). This combination ensures reliable reflectance measurements across a broad wavelength spectrum.

Figure S2 presents the reflectance of the TPV cells obtained by merging the data from both setups previously mentioned, along with data reported in the literature for similar devices using spectrophotometers [1], [2]. For reference, the emission spectrum of the graphite emitter at the maximum operating temperature employed in the TPV experiments is shown.

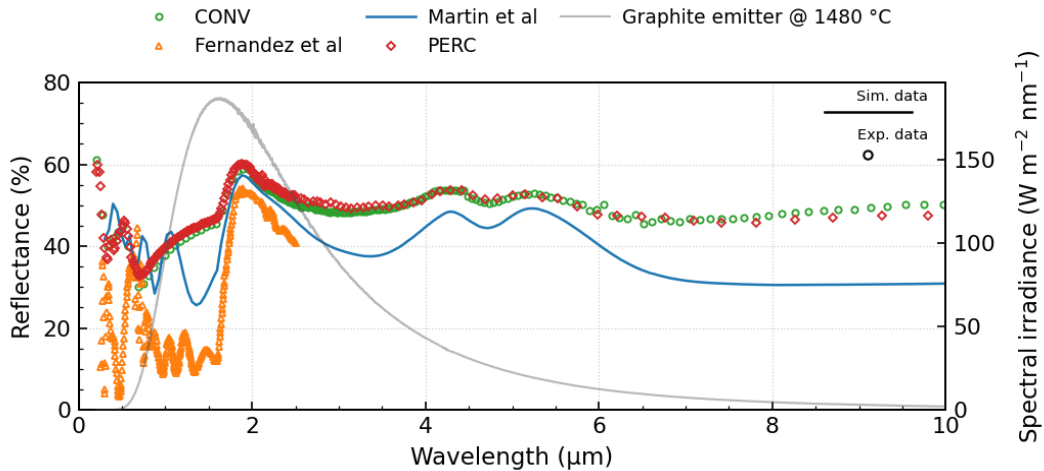

*Figure S2 Reflectivity spectra of the Ge-based TPV devices measured using two different instruments: an FTIR spectrometer equipped with an integrating sphere and a UV–Vis–NIR spectrophotometer. For each individual device (CONV and PERC), the spectra obtained with both instruments show an almost perfect overlap in the common spectral range, with deviations below 1% between 2.2 and 2.4 µm, confirming the excellent reproducibility of the measurements. For comparison, experimental reflectivity data from Fernández et al. [2], [3] are shown up to 2.5 µm, along with the calculated reflectivity presented from Martin et al.[1]. The emission spectrum of the graphite emitter at its maximum recorded temperature is also included for reference.*

Both CONV and PERC devices developed in this work exhibit nearly identical spectral reflectivity, with only a slight improvement for the PERC device at wavelengths below 4 µm. This behavior arises because the effectiveness of the dielectric rear mirror is ultimately limited by strong FCA in the highly doped substrate [1], [4]. As a consequence, the optical benefit provided by the dielectric mirror is reduced to less than a 3 % of absolute value and confined to the short-wavelength region (<4 µm).

Similar FCA-limited behavior is also observed in the devices reported by Martín et al. [1] and Fernández et al. [2], as shown in Figure S2. In the latter case, the incorporation of an anti-reflection coating (ARC) significantly enhances absorption in the IB region (<1.8 µm), at the expense of reduced reflectivity in the OOB region. In both studies, oscillations within the IB region are evident, originating from Fabry–Pérot cavity effects associated with the GaAs layer in Martín's structure and with the combined action of the ARC and GaInP layers in Fernández's device.

Fernández et al. report experimental reflectivity measurements of the complete device only up to 2.5 µm, without extending the spectrum through optical simulations. In contrast, Martín et al. measured the reflectivity of two separate surfaces: the active area without metal and a fully metallized surface using the front-electrode stack [1]. These measurements, also limited to 2.5 μm, were subsequently fitted using a transfer-matrix method (TMM) to extrapolate the optical response beyond the measured range. The reflectivity of the complete device was then reconstructed as a weighted combination of the experimentally measured metallized reflectivity and the TMM-adjusted response of the active area, accounting for the device shadowing factor (SF). This reconstructed reflectivity is the one reported in [1] and Figure S2, and is calculated using the following expression:

$$R_{device} = R_{metal} \cdot SF + R_{Active\ area} \cdot (1 - SF) \qquad (1)$$

To complete the optical description required for the TPV analysis, the spectral emissivity of the thermal emitters considered in this work is examined next. Figure S3 shows the spectral emissivity, i.e. $1 - R_\lambda$, of the two emitters presented and analyzed in this work: 1) the graphite emitter used during the experiments, measured using both FTIR and spectrophotometer (as mentioned above), and 2) the calculated AlN/W spectrally selective emitter presented in [5] and used in [1] for TPV efficiency predictions.

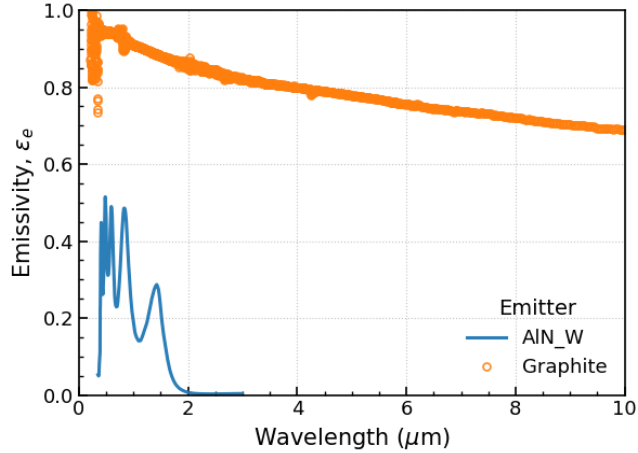

*Figure S3 Emissivity spectra, calculated using room temperature reflectivity data as $1 - R$, of the graphite emitter used in our TPV setup, compared with the emissivity simulated for a AlN/W emitter and used in Martin et al. device [1].*

With the spectral properties of both the TPV devices and the thermal emitters established, we next assess their combined impact on the spectral efficiency of the system. Figure S4 shows the spectral efficiency (SE) as a function of emitter temperature for the CONV, PERC, and Martín *et al.* devices, evaluated using the two emitters described in Figure S3. Two emitter scenarios are considered: a broadband graphite emitter and an idealized spectrally selective AlN/W emitter. The SE is defined as the fraction of the absorbed radiative power lying above the TPV cell bandgap wavelength, $\lambda_g$, thereby quantifying the effect of emitter spectral selectivity on the usable radiation, following the definition introduced in [6]:

$$SE = \frac{\int_0^{\lambda_g} \varepsilon_{eff}\, P_{BB}(T_e, \lambda) d\lambda}{\int_0^{\infty} \varepsilon_{eff}\, P_{BB}(T_e, \lambda) d\lambda} \qquad (2)$$

Here, $\lambda_g$ is the TPV cell bandgap wavelength, $P_{BB}(T_e, \lambda)$ denotes the blackbody spectral power density of the emitter at temperature $T_e$, and $\varepsilon_{eff}(\lambda)$ accounts for the effective radiative coupling between the emitter and the cell.

For both emitter types, all devices exhibit nearly identical SE trends, highlighting that the achievable spectral efficiency is largely constrained by FCA absorption in the highly doped Ge substrate rather than by the rear-side optical design. The selective emitter significantly enhances SE compared to the graphite case, particularly at high emitter temperatures.

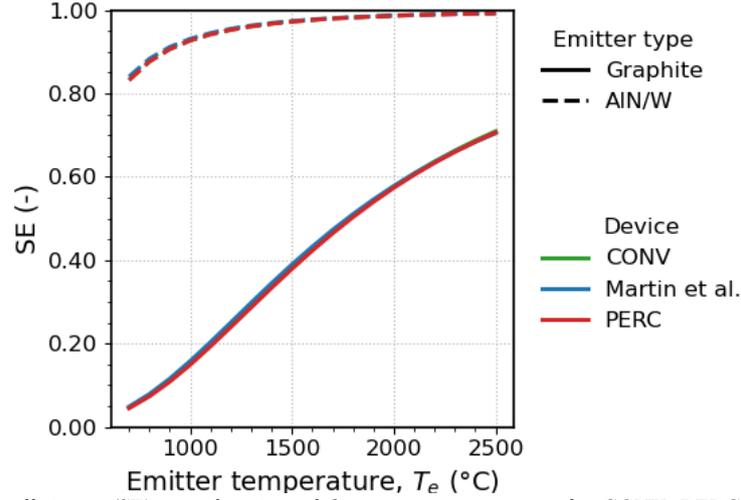

*Figure S4 Spectral efficiency (SE) as a function of the emitter temperature for CONV, PERC, and Martin et al. devices, calculated for two different emitter scenarios: a broadband graphite emitter (solid lines) and a spectrally selective AlN/W emitter (dashed lines). Colors indicate the device architecture.*

### S3. Quantum efficiency (QE)

The internal quantum efficiency (IQE) of the CONV device shown in Figure S5 is obtained through EQE and in-situ reflectivity measurements on a dedicated finger-free test cell, fabricated from the same wafer as the final CONV device later characterized under TPV conditions. These measurements were carried out using a custom-built EQE setup at IES-UPM, based on a W lamp coupled with a grating monochromator and lock-in detection, with a reference Ge photodetector used to monitor and correct lamp intensity fluctuations, the same one used in [1].

Subsequently, the global reflectivity of the final CONV device, including the front metal grid, was measured using an integrating sphere. The EQE of the CONV device was then reconstructed by combining the IQE obtained from the finger-free device with the global reflectivity of the gridded device, according to

$$EQE(\lambda) = IQE(\lambda)\,(1 - R(\lambda)) \qquad (3)$$

where $R(\lambda)$ is the global reflectivity of the final (gridded) CONV device.

For the PERC device, a different methodology was required because a finger-free dedicated cell was not available. In this case, EQE measurements were performed directly on the final (gridded) PERC cell using a commercial Oriel IQE 200 system, covering a spectral range from 300 to 1800 nm. Notably, these EQE measurements were acquired at a single angle of incidence, and the presence of (thick) front-side metal fingers is expected to significantly influence the measured response. To account for this effect, a correction offset was introduced. This offset was determined by comparing the EQE of the CONV device reconstructed using the method described above with the corresponding EQE measured directly using the Oriel system. The resulting offset was then applied to the Oriel-measured EQE of the PERC device to obtain its equivalent EQE. Finally, combining this reconstructed EQE with the integrating-sphere reflectivity of the PERC device enabled the determination of the corresponding IQE of the PERC structure, which is also presented in Figure S5.

It is worth noting that these reconstructed IQE curves will be used only for the TPV model, and therefore, will only impact on the recalculation of the emitter temperatures used in the TPV efficiency analysis. Consequently, they do not influence on the reported experimental efficiencies in this work.

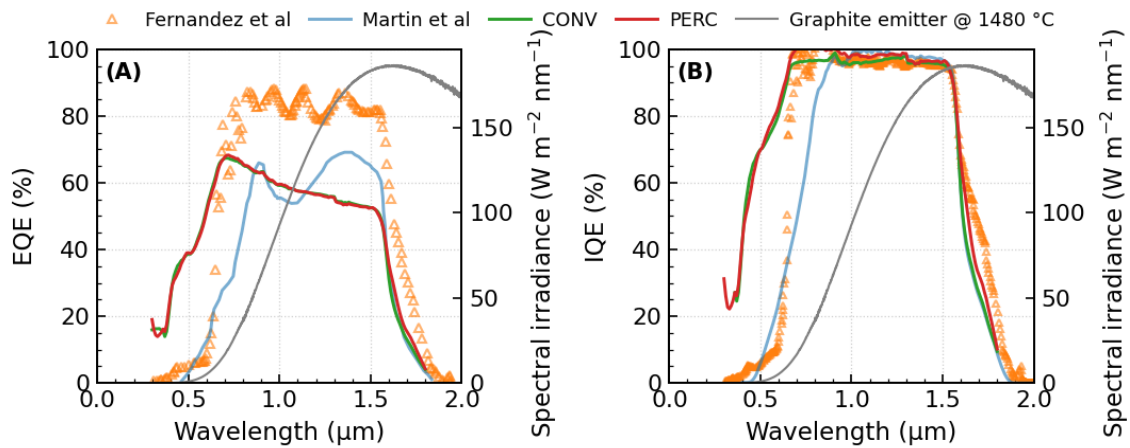

*Figure S5 The plot compares the spectral response of our Ge-based TPV devices, showing both the reconstructed external quantum efficiency (EQE) of each cell and their internal quantum efficiency (IQE). Overlaid as a dashed curve is the spectral irradiance of the graphite emitter at 1496 °C, providing context for the TPV operating spectrum.*

As shown in Figure S5, both CONV and PERC devices exhibit nearly identical EQE spectra, consistent with their identical front-side structures and differing only in rear passivation. Minor deviations observed above 1600 nm may be associated with differences in the passivation stack; however, the strong FCA discussed in the main text largely mask such effects. Due to this strong FCA, potential enhancements arising from the rear structure are not directly visible in the EQE, explaining the nearly identical response of both devices.

When compared with the state of the art, Figure S5A shows that Fernández et al. [2] reported significantly higher EQE, primarily due to the presence of an ARC. Moreover, our structures do not present the Fabry–Pérot oscillations reported by Martin et al. [1], which originate from the presence of a GaAs window layer in their structures. The elimination of the GaAs window layer in our devices enhances the EQE at shorter wavelengths (below 0.8 µm). While this spectral region is less relevant for operation at moderate/low emitter temperatures (e.g. < 1500 °C), it may become advantageous at higher emitter temperatures. For wavelengths above 0.8 µm, our EQE is lower than that reported by Martin et al. [1], mainly due to the higher SF and the absence of the GaAs window layer which behaves as an antireflecting coating, tuning the big EQE oscillation around 1.2 µm. Therefore, in both cases, the higher EQE of previous state of the art devices arises either from a significantly lower SF (6% versus 19% in our devices) or from the presence of an ARC.

It should be noted that the reported experimental data from Martin et al. in [1] correspond to a device characterized without front metal shading. To enable a fair comparison at the device level, the EQE plotted in Figure S5A is calculated as:

$$EQE_{device}(\lambda) = EQE_{active\ area}(\lambda) \cdot (1 - SF) \tag{4}$$

where EQE$_{active}$ refers to the metal-free active-area and EQE$_{device}$ to the device with the corresponding SF.

### S4. Dark I-V characteristics

The dark I-V characteristics of the two devices under study are presented in Figure S6. Each I-V curve was fitted using the single-diode model at 25 °C to extract key parameters: the saturation current density ($J_0$), ideality factor ($n$), and shunt resistance ($R_{sh}$). The series resistance ($R_s$) extracted from the dark IV fitting was not used in the TPV cell model. This is because the current paths in dark conditions are not equal to those in illumination induced by intense irradiance. Instead, the series resistance used in the model was derived from the fitting of the illuminated *I-V* curves obtained in the flash measurements [7], following the method described in [8], which better represent the real behaviour of the devices under high-power operation. All extracted parameters are summarized in Table S1 .

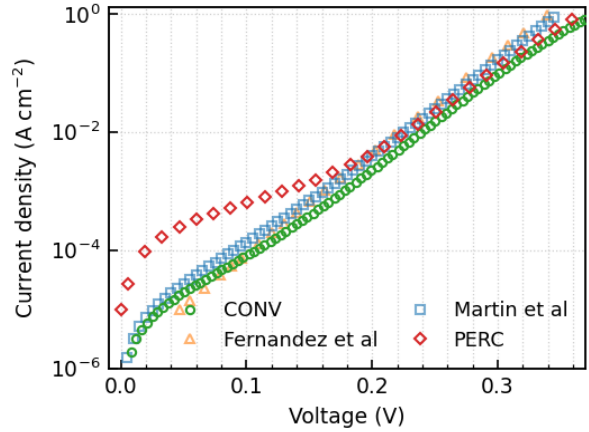

*Figure S6 Dark IV response of the devices reported*

A notably low shunt resistance was observed in the PERC device. This behavior is attributed to the removal of the GaAs contact layer after the mesa etching process, which induces a lateral etch in the perimeter of the cell that creates leakage current paths. This hypothesis is supported by the fact that initial electrical measurements performed prior to this etch yielded shunt resistance values above 10 kΩ·cm², two orders of magnitude greater than the value recorded after the etch (Table S1). Therefore, low shunt resistances can be simply avoided by removing the GaAs layer before the mesa etch.

Nevertheless, this limitation becomes less critical under actual operating conditions. Given the high current densities expected during TPV operation, the impact of a low shunt resistance is significantly mitigated. In such regimes, the current flowing through the shunt path is negligible compared to the total photocurrent, thereby minimizing its effect on overall device performance.

*Table S1 Results of the fitting procedure applied to the reported experimental data.*

|  | Cell area (cm²) [i] | $J_0$ [A/cm²] | n | $R_{sh}$ [$\Omega cm^2$] | $R_s$ [$m\Omega cm^2$] |
|---|---|---|---|---|---|
| *Fernandez et al* [2] | 1.55[ii] | 1.8e-6 | 1 | - | 23 |
| *Martin et al* [1] | 0.096 | 1.6e-6 | 1 | - | 1 |
| *CONV* | 1.05 | 1.83e-06 | 1 | >100 | 7.67 |
| *PERC* | 1.05 | 1.85e-06 | 1 | 193.5 | 17.69 |

[i] *Defined as area inside the mesa.*

[ii] *Only the active area is reported, excluding the busbars.*

## S5. I-V characteristics under flash test conditions

The TPV cells have been characterized using the constant-voltage *I-V* curve flash tester HELIOS 3198 [9]. In this method, the cells are biased at a fixed voltage and exposed to multiple light pulses, allowing the system to record the current generated during the flash decay. By repeating the procedure at different bias voltages, a family of *I–V* curves corresponding to various irradiance levels is obtained. This system thus enables the acquisition of *I–V* curves under variable irradiance conditions.

Figure S7 shows the measured *J–V* curves together with their corresponding fits to the single-diode circuit model. In this figure, the area used for converting current (*I*) to current density (*J*) is 0.85 cm², corresponding to the active area including the metal fingers but excluding the busbars. A simultaneous fitting of all *J–V* curves, obtained under different irradiance conditions, was performed using a multi-curve fitting approach described in [8]. This method captures the device response across a broad range of current densities and voltages, ensuring that the extracted parameters more accurately represent the actual operational regime compared to single-point or low-irradiance fittings.

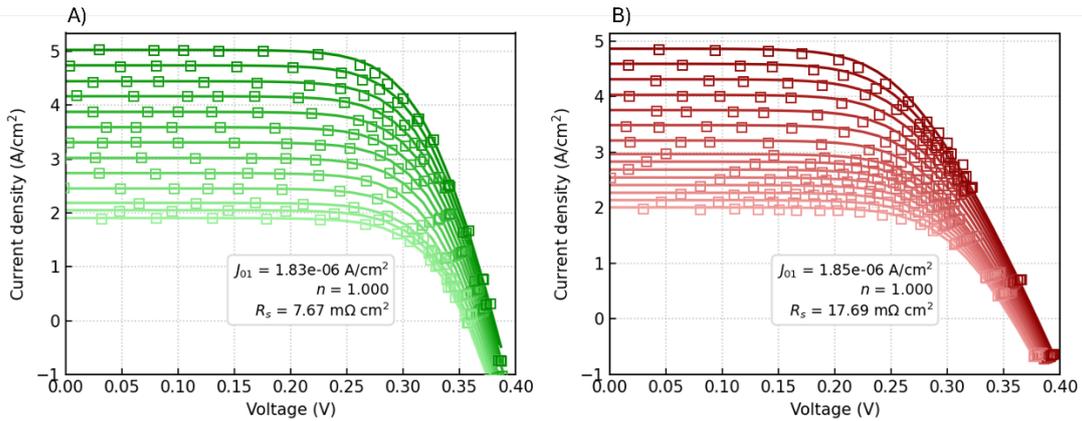

*Figure S7 Fitting of multiple illuminated JV curves for the CONV (left) and PERC (right) devices using the flash setup. The current density was normalized to the active area excluding the busbar, giving an active area of 0.85 cm².*

Notably, the extracted $J_0$ from this fitting (shown in the inset of the figure) is consistent with the value obtained from dark *J–V* curve fitting (Section S4), thereby reinforcing the validity of the results.

Figure S8 presents the key parameters $V_{oc}$ versus $J_{sc}$ (panel A) and FF versus $J_{sc}$ (panel B), extracted from the results in Figure S7, for the two cells developed in this work, together with previously reported data by Fernandez et. al. [2], [3] and Martin et. al. [1], [2]. The figure also includes the results of the single-diode model obtained from the fitted parameters.

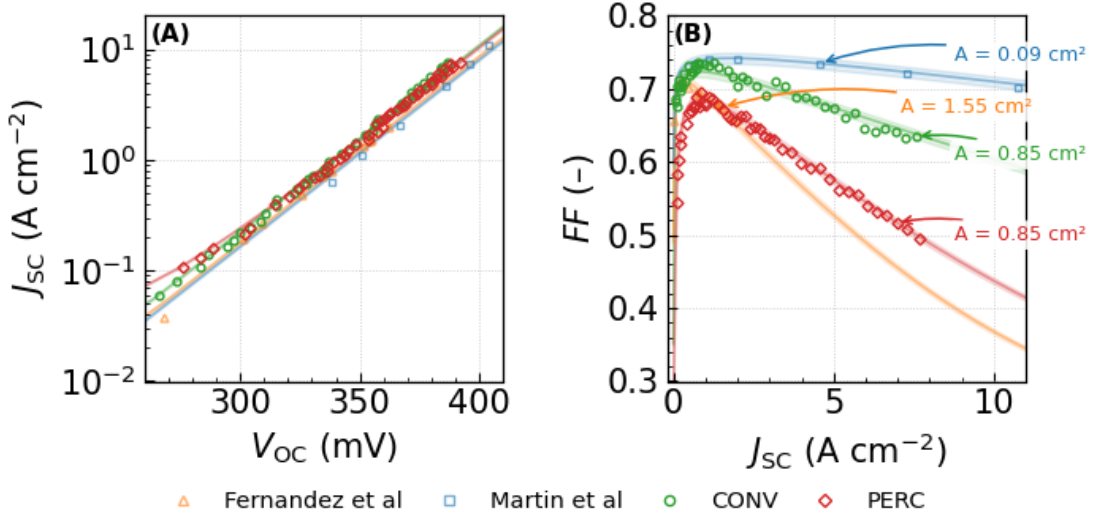

*Figure S8 High-irradiance characterization of our TPV cells under flash-test or laser illumination in the case of Martin et al (non-TPV conditions). (A) $J_{sc}$ versus $V_{oc}$ with single-diode dark-IV fits, confirming the superposition principle. (B) Fill factor versus $J_{sc}$, overlaid with single-diode model predictions computed at 25 °C using device-specific $R_s$, $R_{sh}$, and $J_0$; shaded bands denote ±1%. The data noted by the arrows corresponds to the active area excluding the busbars.*

### S6. Calorimetric characterization and model fitting of TPV cells

This section presents the experimental results obtained from the calorimetric setup [10], which was used to measure the electrical and thermal power outputs of the TPV cells under operating conditions, together with the corresponding model results and fit parameters.

*Table S2 and Table S3 Summary of PERC data in the experimental TPV setup. The temperature values in brackets correspond to the experimental measurements obtained through our TPV setup, while the others represent the corrected values using the method described in Section 2.3. Same considerations as explained in Table S3.*

Table S3 summarize the relevant parameters (temperature, $I_{sc}$, $V_{oc}$, $I_{mpp}$, $V_{mpp}$) for the CONV and PERC cells, respectively, along with their output power ($P_{mpp}$) and dissipated heat ($Q_{dis}$). These values are then used to determine the TPV conversion efficiency through the following equation:

$$\eta_{TPV} = \frac{P_{mpp}}{P_{mpp} + Q_{dis}}$$

(5)

Both uncorrected values (reported in brackets) and corrected values for the emitter and cell temperatures are included. As explained in Section 2.3, the temperature correction relies on two factors, $r_e$ and $r_c$, introduced to improve the agreement between the experimental results and the model. The optimal correction factors that provide the best fit are summarized in Table S4.

In addition, Table S2 and Table S3 report the saturation current density, $J_0$, obtained from fitting the single-diode model to the $I$–$V$ curves measured under different TPV operating conditions (emitter temperatures). The fitted curves are shown in Figure S9 for the CONV (panel A) and PERC (panel B) cells. A clear increase in $J_0$ with increasing emitter temperature is observed, which is attributed to the corresponding rise in cell temperature.

*Table S2 Summary of CONV data measured under the experimental TPV setup. Temperatures shown in brackets correspond to the values directly measured by a thermocouple located in the calorimeter, a few millimeters away from the cell. Thermal grease is used between the copper components of the setup, while a thermally conductive epoxy ensures the bonding between the cell and the copper pedestal. The remaining temperatures correspond to corrected values obtained using the procedure described in Section 2.3, which accounts for thermal resistances in the mounting stack and for temperature gradients across the Ge substrate, leading to a higher effective p–n junction temperature and influencing the extracted $J_0$.*

| T emitter [°C] | T cell [°C] | $I_{sc}$ [A] | $V_{oc}$ [mV] | $I_{mpp}$ [A] | $V_{mpp}$ [mV] | $P_{mpp}$ [W] | TPV efficiency [%] | Q [W] | FF [%] | $J_0$ [A/cm2] |
|---|---|---|---|---|---|---|---|---|---|---|
| 956 (980) | 27.7 (26.9) | 0.25 | 307.94 | 0.23 | 243.93 | 0.06 | 1.90 | 2.84 | 71.62 | 1.72E-06 |
| 1041 (1073) | 27.9 (26.9) | 0.42 | 321.16 | 0.39 | 253.19 | 0.10 | 2.66 | 3.64 | 72.36 | 1.75E-06 |
| 1113 (1155) | 28.1 (26.9) | 0.65 | 331.22 | 0.60 | 262.56 | 0.16 | 3.41 | 4.47 | 72.79 | 1.79E-06 |
| 1211 (1267) | 28.5 (26.9) | 1.08 | 342.50 | 1.00 | 271.77 | 0.27 | 4.44 | 5.82 | 72.83 | 1.91E-06 |
| 1293 (1366) | 29.0 (27.1) | 1.61 | 350.61 | 1.45 | 281.14 | 0.41 | 5.36 | 7.22 | 72.48 | 2.07E-06 |
| 1364 (1455) | 29.5 (27.2) | 2.21 | 356.89 | 2.02 | 281.08 | 0.57 | 6.13 | 8.67 | 71.70 | 2.25E-06 |
| 1425 (1535) | 29.8 (27.1) | 2.88 | 361.73 | 2.61 | 281.01 | 0.73 | 6.74 | 10.15 | 70.64 | 2.44E-06 |
| 1462 (1586) | 30.4 (27.4) | 3.38 | 363.89 | 3.00 | 285.14 | 0.86 | 7.16 | 11.18 | 69.65 | 2.62E-06 |
| 1480 (1610) | 31.1 (27.9) | 3.64 | 364.14 | 3.32 | 275.79 | 0.92 | 7.29 | 11.70 | 69.11 | 2.83E-06 |

*Table S3 Summary of PERC data in the experimental TPV setup. The temperature values in brackets correspond to the experimental measurements obtained through our TPV setup, while the others represent the corrected values using the method described in Section 2.3. Same considerations as explained in Table S3.*

| T emitter [°C] | T cell [°C] | $I_{sc}$ [A] | $V_{oc}$ [mV] | $I_{mpp}$ [A] | $V_{mpp}$ [mV] | $P_{mpp}$ [W] | TPV efficiency [%] | Q [W] | FF [%] | $J_0$ [A/cm2] |
|---|---|---|---|---|---|---|---|---|---|---|
| 920 (952) | 25.2 (24.8) | 0.18 | 296.64 | 0.16 | 230.75 | 0.04 | 1.89 | 1.92 | 68.79 | 1.79E-06 |
| 1002 (1045) | 25.4 (24.9) | 0.32 | 308.57 | 0.28 | 244.02 | 0.07 | 2.61 | 2.55 | 69.95 | 1.90E-06 |
| 1073 (1128) | 25.7 (25.0) | 0.49 | 316.99 | 0.43 | 251.82 | 0.11 | 3.27 | 3.23 | 70.24 | 2.13E-06 |
| 1153 (1226) | 26.0 (25.1) | 0.78 | 324.45 | 0.70 | 251.83 | 0.18 | 4.04 | 4.17 | 69.57 | 2.53E-06 |
| 1238 (1336) | 26.2 (25.1) | 1.23 | 330.17 | 1.11 | 251.44 | 0.28 | 4.87 | 5.41 | 68.39 | 3.21E-06 |
| 1309 (1434) | 26.6 (25.2) | 1.76 | 332.78 | 1.56 | 251.43 | 0.39 | 5.52 | 6.70 | 66.95 | 4.16E-06 |
| 1372 (1528) | 27.1 (25.4) | 2.39 | 333.20 | 2.12 | 243.99 | 0.52 | 5.98 | 8.11 | 64.68 | 5.59E-06 |
| 1426 (1617) | 27.6 (25.6) | 3.13 | 331.76 | 2.74 | 236.60 | 0.65 | 6.30 | 9.63 | 62.33 | 7.81E-06 |

*Table S4 Fitting parameters for the correction of temperatures measured in the TPV calorimetric set up.*

| Device | $r_c$ [°C/W] | $r_e$ [°C/W] |
|---|---|---|
| CONV | 0.27 | 4.06 |
| PERC | 0.22 | 5.67 |

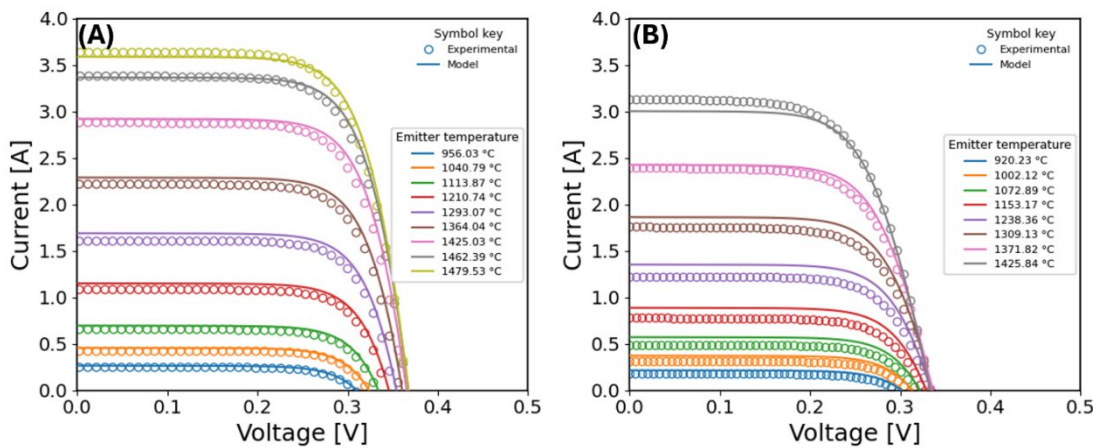

## S7. Electrodes design and series resistance analysis

This section analyses the contribution of the different resistivity sources to the total series resistance of the device. Three main contributions are considered, namely: the front contact, the bulk and the rear contact:

$$R_s = R_{front} + R_{Bulk} + R_{Rear} \qquad (6)$$

The front contact contribution ($R_{front}$) which is common to both CONV and PERC architectures, is evaluated using the analytical lumped-element model described in [11]. Within this framework, the front resistance can be expressed as the sum of three components and their corresponding contributions:

$$R_{front} = R_{me} + R_{le} + R_{ce} \qquad (7)$$

where the individual terms are given by:

$$R_{me} = \frac{dl^2 \rho_m}{12wb} \qquad (8)$$

$$R_{le} = \frac{d^2 R_\square}{12} \qquad (9)$$

$$R_{ce} = \frac{\rho_{ce}}{\left( w/b \right)} \qquad (10)$$

- $R_{front}$ is the total front-side series resistance contribution.

- $R_{me}$ is the resistance associated with current transport along the front metal grid fingers.

- $R_{le}$ is the lateral carrier transport resistance in the emitter between adjacent grid fingers.

- $R_{ce}$ is the metal–semiconductor contact resistance at the front interface.

The geometrical and material parameters are defined as follows:

- $d$ is the pitch between adjacent front metal fingers.

- $w$ is the width of the front metal fingers.

- $l$ is the effective finger length, defined by the active cell width minus the busbar regions.

- $b$ is the thickness of the front metal fingers.

- $\rho_m$ is the electrical resistivity of the front metal.

- $R_\square$ is the sheet resistance of the emitter layer.

- $\rho_{ce}$ is the specific contact resistivity of the metal–semiconductor interface at the front contact.

A schematic illustration of these contributions and their physical origin is provided in Figure S10.

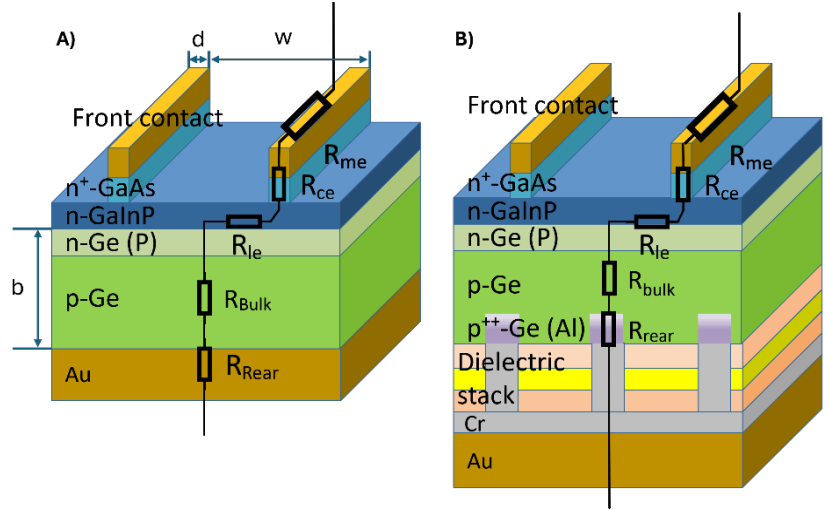

*Figure S10. (A) Schematic cross-section of the CONV device architecture, indicating the relevant geometrical dimensions and the parameters of the front metal grid used in the resistance modeling. (B) Schematic cross-section of the PERC device architecture, showing the same front-grid geometrical parameters together with the additional rear-contact geometry.*

The bulk contribution ($R_{bulk}$) is calculated in a different way for both CONV and PERC devices. For CONV cells, the bulk contribution is simply obtained by multiplying the bulk resistivity ($\rho = 3.36 \times 10^{-2}$ $\Omega$·cm, value extracted from [12], [13]) by the wafer thickness:

$$R_{bulk,CONV} = \rho w \qquad (11)$$

In the case of the PERC device, the calculation must account for a more complex 3D distribution of the current towards the punctual contacts in the rear side. For that, we use the analytical approach described in detail in [14], [15], [16], according to which the bulk contribution of a PERC device is given by:

$$R_{bulk,PERC} = P^2 \cdot \frac{\rho \cdot r}{2} \arctan\left(\frac{2 \cdot w}{r}\right) + \rho \cdot w \cdot \left(1 - e^{-\left(\frac{w}{P}\right)}\right) \qquad (12)$$

where $P$ is the pitch between contact points (arranged in a square pattern), $\rho$ is the substrate resistivity, $r$ is the contact radius, and $w$ is the wafer thickness. The first term represents the spreading resistance ($R_{spread}$) associated with current converging into isolated point contacts in the large-pitch limit. The second term is an exponential interpolation ($R_{int}$) term that captures the transition towards the dense-contact regime ($P \ll w$), in which the rear surface behaves increasingly like a fully metallized back contact.

Finally, the rear contact resistance ($R_{rear}$) is obtained for both PERC and CONV devices such as:

$$R_{rear} = \frac{\rho_c}{f_c} \qquad (13)$$

where $\rho_c$ is the specific contact resistance, and $f_c$ is a contact coverage parameter. This parameter is equal to unity for CONV devices, reflecting full rear-contact coverage, whereas for PERC devices it corresponds to the fraction of laser-processed area, given by $f_c = \pi r^2 / P^2$.

For convenience, we define the LFC resistance of the PERC device as the addition of bulk and rear contact resistances:

$$R_{LFC}[\Omega cm^2] = \frac{\rho_c}{f_c} + R_{bulk,PERC} \qquad (14)$$

Figure S11 summarizes the resistance components required to calculate the total series resistance according to equations (6)–(7). Each quantity is shown in a dedicated panel: the electroplated front-metal resistivity $\rho_m$ (panel A), the emitter sheet resistance $R_\square$ (panel B), the front metal–semiconductor specific contact resistivity $\rho_{ce}$ (panel C), and the rear contact resistance $R_{rear}$ (panel D). The reported values were extracted from Transfer Line Method (TLM) measurements (B, C and D) and metal-layer sheet resistance characterization (A). Measurements were performed on multiple devices, leading to the statistical dispersion reflected in the box plots. These experimentally determined parameters are subsequently used as inputs for the series-resistance modelling of both CONV and PERC architectures.

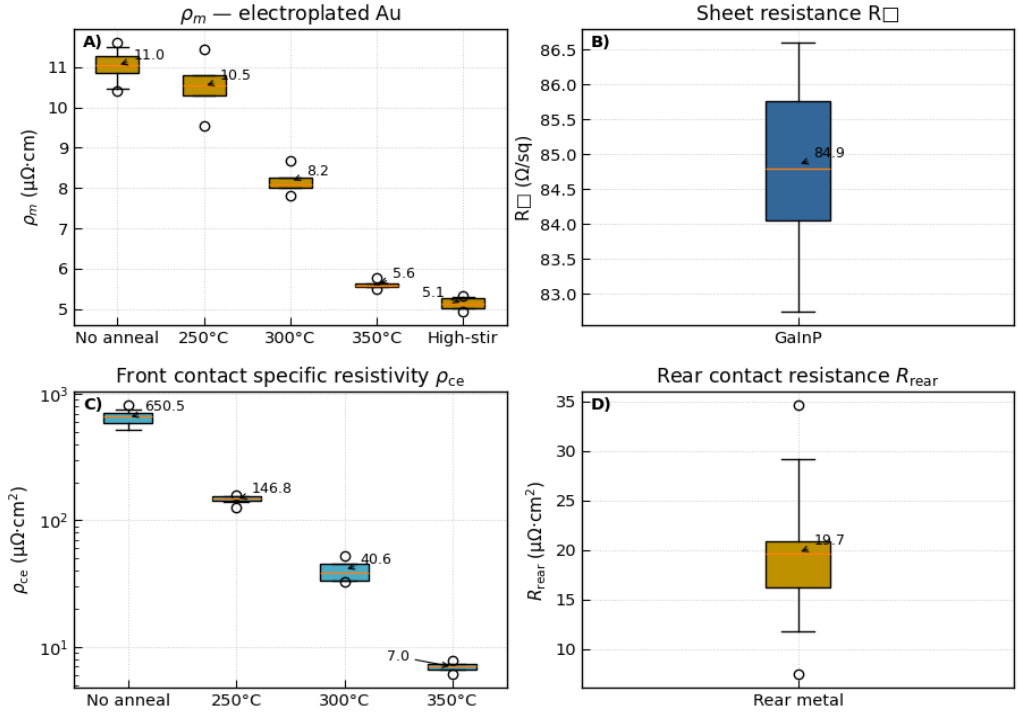

*Figure S11. (A) Electrical resistivity of electroplated Au as a function of annealing temperature and stirring conditions. (B) Sheet resistance of the GaInP window layer. (C) Specific contact resistivity $\rho_{ce}$ of the front metal contact as a function of annealing temperature. (D) Rear contact resistance of Au contacts on $1 \times 10^{17}\ cm^{-3}$ doped Ge substrates.*

The specific contact resistivity ($\rho_c$) of the rear contact on for the PERC devices was not directly measured experimentally. Instead, it was inferred from the total series resistance measured for the PERC device and the calculated front contact contribution (obtained as the sum of its individual components using the experimentally determined parameters reported). Subtracting the calculated front contribution from the total series resistance yields a value of 3.5 mΩ·cm² associated with the rear contact and substrate, which corresponds to a $\rho_c$ of $5.1 \cdot 10^{-5}\ \Omega \cdot cm^2$, following equation (14). This value is in line with the $\rho_c$ of $5.7 \cdot 10^{-5}\ \Omega \cdot cm^2$ reported in [15] for laser-fired point contacts using aluminum as the contact metal.

Contact design and series resistance of optimized devices:

Using the empirical values for each resistive contribution summarized previously and the corresponding analytical expressions, the geometrical parameters of the front and rear contacts were optimized. Figure S12A shows a two-dimensional map of the front-grid contribution to the series resistance as a function of finger width $w$ and finger pitch $d$. This map was used to identify geometries that minimize $R_{front}$ while maintaining a practical shading factor (SF = $w/d$), thereby locating an equilibrium between electrical optimum and fabrication constraints. The experimental point on the graphs corresponds to the fabricated CONV device for reference.

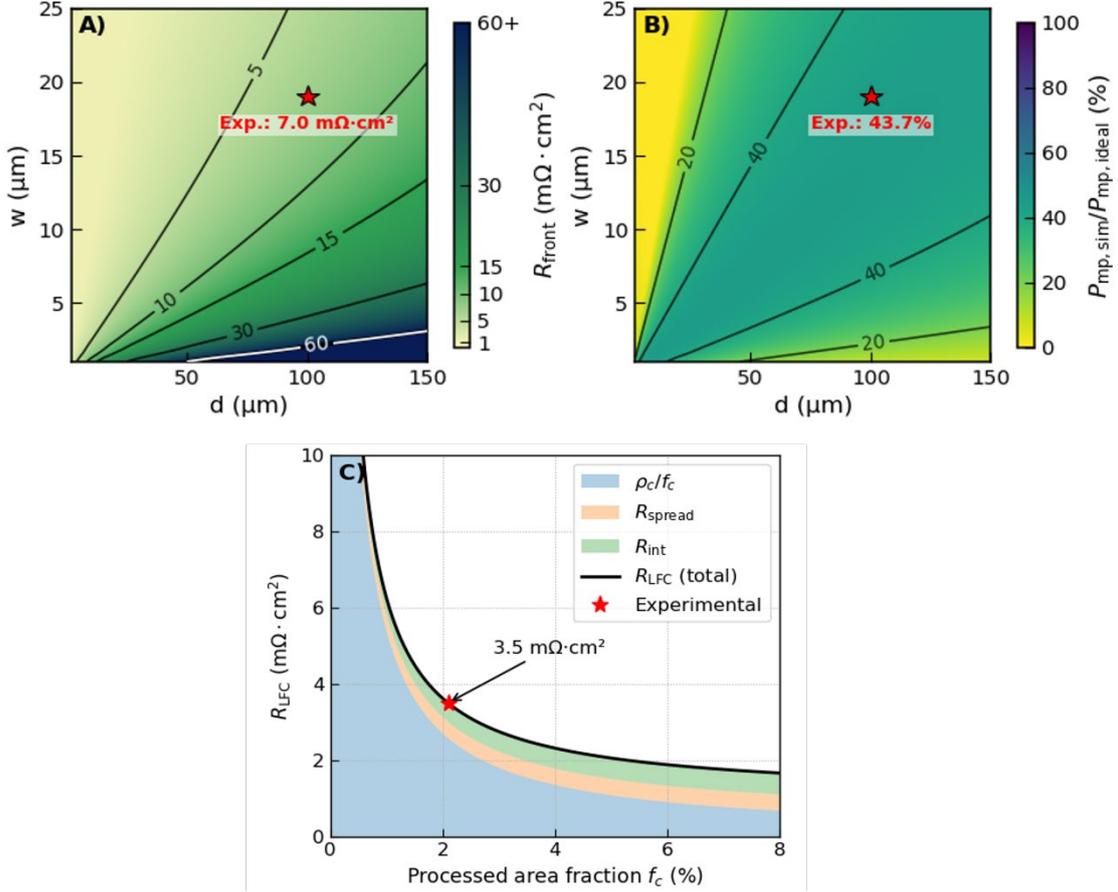

*Figure S12. A) Two-dimensional map of the front-grid contribution to the series resistance as a function of finger width $w$ and finger pitch $d$, including iso-resistance contours. B) Corresponding map of the simulated electrical power ratio relative to the ideal case (Rs=0, SF=0), illustrating the impact of front-grid geometry on electrical performance. In both A) and B), the experimental design point for CONV devices is indicated for reference. C) Rear laser-fired contact resistance $R_{LFC}$ as a function of the processed rear area fraction $f_c$, obtained from the analytical model. The total resistance is shown together with the separate contributions from the contact term $\rho_c/f_c$, the spreading resistance $R_{spread}$, and the interaction term $R_{int}$, as defined in Eqs. (10) and (12). The experimental result is highlighted for reference.*

Figure S12B translates the same $(d, w)$ design space into a performance-oriented metric by reporting the simulated maximum-power ratio relative to an ideal reference case with SF = 0 and $R_s$ = 0 at the case of an emitting temperature of 1500 ºC. The simulations include the geometry-dependent front-grid series resistance together with fixed bulk and rear-contact contributions common to all $(d, w)$ configurations. Shading was treated spectrally by combining the measured reflectivity of the Au front metal with the measured reflectivity of the CONV device (SF ≈ 19 %) to reconstruct the active-area response and derive a geometry-dependent EQE (and therefore $J_{sc}$) for each $(d, w)$ pair. In this way, Figure S12B highlights the trade-off between reducing the front-grid series resistance (favoring denser and/or wider grids) and limiting shading losses (favoring sparser and/or narrower grids), enabling the identification of grid geometries that maximize deliverable power under TPV operation.

Importantly, this optimization does not yield a single unique solution, but rather a broad region of grid geometry providing comparable electrical performance. This tolerance allows practical fabrication considerations to guide the final choice. The systematic widening of the fingers observed after fabrication—caused by photoresist profile evolution and metal electroplating—is an intrinsic feature of the process and occurs for all devices, independently of the nominal mask design. As a result, although the photomask defined a nominal finger width of 15 µm, the effective width in the fabricated devices is approximately 19 µm. This effect must be considered in the design process and find a way to minimize it during fabrication.

Figure S12C shows the $R_{LFC}$ contribution of the PERC device as a function of the processed area fraction of $f_c$. Increasing $f_c$ reduces $R_{LFC}$ by shortening the current-spreading path; however, it simultaneously increases the disruption of the rear mirror, leading to a degradation of its optical reflectivity and, consequently, of the overall optical performance. This trade-off between electrical resistance and rear-side reflectivity has been quantitatively analyzed in previous studies on laser-fired contacts, where an optimal contact pitch was identified to balance both effects [15]. Following this established approach, an intermediate processed fraction of approximately 2.1 %, corresponding a pitch of P≈60 µm, is selected in this work, providing sufficiently low series resistance while largely preserving the optical quality of the rear mirror.

Taken together, the analyses presented in Figure S13 enable the experimental results to be interpreted in terms of contact geometry and scaling effects. The resistance model assembled from the experimentally extracted contributions shows good agreement with the measured values across different devices, supporting its use as a consistent framework for the analysis that follows. The Figure S13 summarizes how front-grid geometry, rear-contact processing, and device scaling translate into the total series resistance for different device architectures, assuming a two-busbar configuration with fixed busbar dimensions and edge-to-contact tolerances. Importantly, when increasing the device area, the front-grid layout is not re-optimized. Instead, key geometrical and material parameters, such as finger width, finger pitch, metal thickness, busbar geometry, and contact resistivities, are kept constant. This approach focuses on the effect of device scaling on the series-resistance budget, without introducing additional grid re-optimization effects.

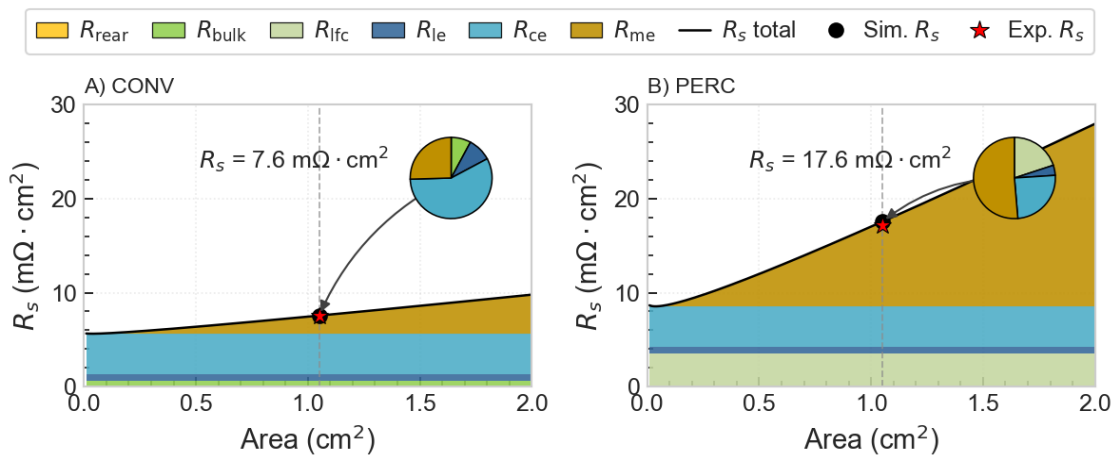

*Figure S13 (A) CONV device as fabricated (C) PERC device as fabricated.*

As observed, in the as-fabricated device in Figure S13 the bottleneck of the series resistance is located at the front contact, arising either from the $R_{ce}$ or from the finite $R_{me}$. The rear contact contribution, despite it being less relevant in both devices, is more significant in PERC devices. This is mostly attributed to the lower $f_c$, which makes the impact of $\rho_c$ more noticeable.

The relatively high front contact resistance in both PERC and CONV devices ($R_{ce}$= $6.4 \times 10^{-4}$ $\Omega \cdot cm^2$) originates from the unannealed metal (AuGe/Ni/Au)/semiconductor interface. Although this contact resistance can, in principle, be further reduced through post-deposition annealing, all annealing attempts resulted in short-circuiting devices. This behavior is attributed to the thin GaAs contact layer, which is insufficient to prevent metal spike penetration into the p–n junction during annealing. Such penetration ultimately leads to severe leakage current paths and device failure.

The $R_{me}$ greatly depends on the conductivity of the electroplated gold layer, which is primarily determined by the electroplating conditions. For instance, increasing electrolyte agitation during deposition promotes a denser and more uniform gold layer, thereby improving its conductivity and reducing resistive losses. The degree of agitation depends on the stirrer and the configuration of the plating setup; the key indicator of proper mixing is the formation of a visible vortex in the electrolyte. Because of the fabrication sequence and the learning curve associated with the process, this optimization was not implemented in the present PERC device, which explains the higher series resistance of this device. However, it represents a viable strategy to enhance device performance.

Projecting series resistance for scaled up and scaled down scenarios:

According to the previous discussion there are three main strategies for reducing the front-side contact series resistance: (i) annealing to improve the front specific contact resistivity, (ii) increasing the thickness of the GaAs contact layer to prevent short-circuiting of the pn junction, and (iii) optimizing the electroplating process to enhance gold conductivity. Based on this, Figure S14 presents a comparative analysis of the series resistance as a function of the area for both device architectures, along with a scaled-down configuration to reproduce the work of Martin et al [1].

In this figure, panels A, B, and C show the contribution of the series resistance Rs as a function of the total device area, assuming a square layout and keeping the finger width ($w$) at 15 $\mu$m and the spacing ($d$) at 100 $\mu$m. This configuration represents the design choice for the scaled-up device (total device area of 1.05 cm²). In this analysis, the total device area is used as the scaling parameter on the horizontal axis, while the busbar dimensions and edge margins are kept constant. Consequently, the electrically active area is slightly smaller than the total device area but scales proportionally with it.

In panels D, E, and F, the cell remains square, but the finger width is reduced to 6 $\mu$m and the spacing to 94 $\mu$m, representing the design choice for the scaled-down configuration (total device area of 0.096 cm²). Using the total device area as the scaling parameter ensures a consistent comparison of the resistive contributions across device sizes while preserving realistic contact layouts and fixed peripheral regions.

The Figure S14 shows that, once the front contact is improved through annealing (e.g. panel C), the bottleneck in PERC devices shifts to the rear side of the cell, mainly due to the relatively small, contacted area.

These calculations are subsequently employed in Section 4.1 of the main text, where the TPV model described above is applied to estimate the projected conversion efficiencies under two representative operating scenarios. The first scenario considers large-area devices operating with non-selective emitters, such as graphite, which are relevant for practical high-power implementations. The second scenario focuses on smaller devices combined with spectrally selective emitters, where enhanced photon management can be exploited to improve efficiency. Building on these two cases, the analysis is further extended to explore potential performance improvements by systematically maximizing the achievable efficiency within the constraints imposed by the proposed geometries. In this way, the calculations do not merely reproduce measured behavior but provide a consistent framework to assess the performance limits and optimization pathways of Ge-based TPV devices under realistic and idealized operating conditions.

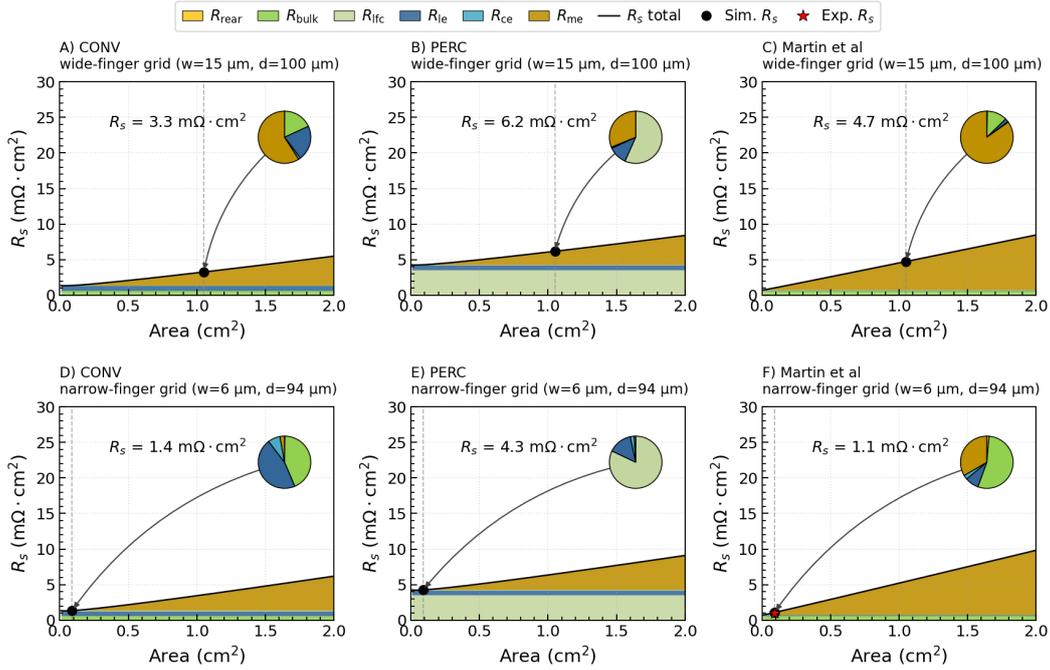

*Figure S14 (A) CONV device after the annealing treatment; (B) CONV device downscaled to the Martin et al. form factor; (C) PERC device after the annealing treatment; (D) PERC device downscaled to the Martin et al. form factor; (E) Martin et al. device scaled to our form factor; and (F) Martin et al. device as reported in [1]. For the rear contact in the PERC devices, the processed area fraction $f_c$ is defined by the laser-fired contact pitch, which is fixed to 60 μm in all cases, corresponding to the value implemented in the fabricated devices.*

## S8. Key Assumptions Behind Reported State-of-the-Art Ge TPV Cell Results

In this section, we provide a detailed analysis of the reported data from two reference works (Fernández *et al.* [2] and Martin *et al.* [1]) that are discussed throughout this study.

The work by Fernández et al. [2], [3] reflects an early semi-empirical approach used by the TPV community to estimate device performance before calorimetric experimental setups became more popular. First, the reflectivity spectrum is experimentally reported only between 0.2 and 2.5 μm, omitting longer wavelengths that are essential for TPV applications. Second, no optical cavity is considered, and the view factor is simply assumed to be 1. Finally, and most importantly, the calculations assume a microstructured wolfram emitter heated to 1100 ºC and with spectral cut-off, i.e. zero emissivity beyond 1.88 μm (the indirect bandgap of germanium). Therefore, the work neglects optical losses in the sub-bandgap region. While useful for providing a quick estimation of the ultimate efficiency potential, this idealization leads to inflated efficiency values that overlook long-wavelength thermal radiation and real optical behavior. The reported TPV efficiency of 16.5 % is calculated analytically using the standard PV efficiency relation:

$$\eta_{TPV} = \frac{I_{SC}V_{OC}FF}{P_{TPV}} \qquad (15)$$

with $I_{sc}$, $V_{oc}$, and FF taken from the I-V curve, and $P_{TPV}$ estimated from the truncated emission spectrum of a micro-structured wolfram emitter, as explained above. The assumption of perfect reflectivity beyond the cut-off wavelength artificially reduces the absorbed power, again leading to an overestimated efficiency figure.

Martin et al. [1] follow a more refined but still semi-empirical methodology, motivated by the same lack of calorimetric TPV measurement setups. The most relevant assumption regards the use of a simulated spectrally selective emitter, inspired from the work of [5] and reproduced using TMM. The emitter emissivity is purely simulated and lacks experimental optical validation at TPV operating temperatures, with its thermal stability inferred from literature data (AlN/W reported to be stable ex situ up to ~1700 °C, while AlN is expected to decompose above ~1500 °C under high-vacuum conditions). Accordingly, the emissivity adopted in [1] should be regarded as an idealized, simulation-based spectrum rather than an experimentally validated emitter response.

Moreover, in Martin et al. [1], the cell reflectance is not obtained from direct global measurements of the final metallized device. Instead, optical characterization is performed on a reference device without front metallization, for which the EQE and reflectance of the active area are measured directly. In a second step, the optical response of the front metal is measured independently on a separate metal film. The device-level optical response is then reconstructed by combining the measured reflectance of the metal-covered regions with that of the optically active semiconductor area using a geometrical model based on the SF. Specifically, the EQE of the device is obtained by scaling the EQE of the active area by the unshaded fraction, following Equation (*4*), while the effective reflectance is calculated as an area-weighted combination of the metal and semiconductor contributions as in Equation (1). This approach enables the reconstruction of the effective EQE and reflectance of the metallized TPV device without direct global optical measurements.

The use of small-area devices for electrical characterization significantly reduces the impact of series resistance, particularly contributions associated with front-grid geometry, as discussed in the previous section. This facilitates electrical measurements under high current densities and allows the device to operate closer to its ideal electrical behavior. However, the resulting electrical performance is not directly representative of larger-area TPV devices operating under practical high-irradiance conditions, where series resistance and front-grid design play a much more critical role.

The overall TPV efficiency is estimated using the oversimplified expression $\eta_{TPV} = P_{mpp}/(VF \cdot (P_{emitted} - P_{reflected}))$, where VF is the view factor, $P_{emitted}$ denotes the emitted spectrum in vacuum, and $P_{reflected}$ is obtained by multiplying $P_{emitted}$ by the cell reflectance. However, this formulation provides only a crude estimate of the net radiative power absorbed by the cell, as it neglects the complex radiative exchange between the emitter and the photovoltaic cell. A more accurate description requires the use of an effective emissivity, as adopted in this work, or alternatively, full ray-tracing simulations. This simplification becomes particularly problematic for emitters with low emissivity—such as the emitter considered in their study—because radiation reflected by the cell is not efficiently reabsorbed but instead undergoes multiple reflections at the emitter, leading to a significant misestimation of the absorbed power.

Overall, both approaches provide estimations of TPV efficiency that rely heavily on important assumptions. These include truncated or simulated spectral data, simplified geometries, and analytical corrections for reflection and shadowing. Such assumptions must be carefully considered when comparing with directly measured efficiencies.